\documentclass[%
 reprint,
 amsmath,amssymb,
 aps,
prx,
]{revtex4-2}
\usepackage{amsmath,amssymb,amsfonts}
\usepackage{algorithmic}
\usepackage{graphicx}
\usepackage{textcomp}
\usepackage{xcolor}
\usepackage[colorlinks=true, hyperindex, breaklinks, linkcolor=blue, urlcolor=blue, citecolor=blue]{hyperref} 
\usepackage[capitalise]{cleveref}
\usepackage{ifthen}
\usepackage[normalem]{ulem}
\usepackage{siunitx}
\usepackage[caption=false]{subfig}
\usepackage{enumerate}
\usepackage[shortlabels]{enumitem}
\usepackage{comment}
\usepackage{tabularray}
\usepackage{multirow}

\newcommand{\tr}{\operatorname{Tr}}

\newcommand{\sx}{$\sqrt{\text{X}} $ }

\newcommand{\samplesize}{1000 }

\newcommand{\mseeloss}{\mathcal{L}_{\text{SE[E]}}}
\newcommand{\maefloss}{\mathcal{L}_{\text{AE[F]}}}
\newcommand{\mseemodel}{\check{f}_{\text{SE[E]}}}
\newcommand{\maefmodel}{\check{f}_{\text{AE[F]}}}

\newboolean{ShowComments}
\setboolean{ShowComments}{true}  
\ifthenelse{\boolean{ShowComments}}%
	{
		\newcommand{\ColorComment}[3]{%
				{\colorbox{#1}{\color{white}   \textsf{\textbf{#2}}} \textcolor{#1}{#3}}}

        \newcommand{\blk}{\color{black}}
        
	}%
	{
		\newcommand{\ColorComment}[3]{}

        \newcommand{\blk}{\color{black}}
        
	}%
\definecolor{michalcolor}{RGB}{255,127,80}
\definecolor{porametcolor}{HTML}{8b5cf6}
\definecolor{areeyacolor}{HTML}{ef4444}
\definecolor{rdvcolor}{HTML}{eab308}

\newcommand{\control}{\mathbf{\Theta}}

\def\BibTeX{{\rm B\kern-.05em{\sc i\kern-.025em b}\kern-.08em
    T\kern-.1667em\lower.7ex\hbox{E}\kern-.125emX}}

\begin{document}

\title{Graybox characterization and calibration with finite-shot estimation on superconducting-qubit experiments
}

\author{Poramet Pathumsoot}
\email{poramet@sfc.wide.ad.jp}
\affiliation{Graduate School of Media and Governance, Keio University Shonan Fujisawa Campus, Kanagawa, Japan}

\author{Areeya Chantasri}
\email{areeya.chn@mahidol.ac.th}
\affiliation{Optical and Quantum Physics Laboratory, Department of Physics, Faculty of Science, Mahidol University, Bangkok 10400, Thailand}

\author{Michal Hajdu\v{s}ek}
\email{michal@sfc.wide.ad.jp}
\affiliation{Graduate School of Media and Governance, Keio University Shonan Fujisawa Campus, Kanagawa, Japan}

\author{Rodney Van Meter}
\email{rdv@sfc.wide.ad.jp}
\affiliation{Faculty of Environment and Information Studies, Keio University Shonan Fujisawa Campus, Kanagawa, Japan}

\begin{abstract}
    Characterization and calibration of quantum devices are necessary steps to achieve fault-tolerant quantum computing. As quantum devices become more sophisticated, it is increasingly essential to rely not only on physics-based models, but also on predictive models with open-loop optimization. Therefore, we choose the Graybox approach, which is composed of an explicit (whitebox) model describing the known dynamics and an implicit (blackbox) model describing the noisy dynamics in the form of a deep neural network, to characterize and calibrate superconducting-qubit devices. By sending a set of selected pulses to the devices and measuring Pauli expectation values, the Graybox approach can train the implicit model and optimize gates based on specified loss functions. We also benchmark our optimized gates on the devices and cross-testing predictive models with two types of loss functions, i.e., the mean squared errors (MSE) of expectation values and the absolute errors (AE) of average gate fidelities (AGF). While the Graybox method allows for flexibility of the implicit noise model, its construction relies on a finite measurement shots dataset. We thus apply the decomposition of expected MSE loss to show that the finite-shot estimation of expectation values is the main contribution to the minimum value achievable of the expected MSE loss. We also show that the expected loss is an upper bound of the expected absolute error of AGF between the exact value and model prediction. Our results provide insights for quantum device characterization and gate optimization in experiments where only finite shots of data are available.
\end{abstract}

\maketitle

\section{Introduction} \label{sec:intro}

High-fidelity quantum gates with error rates below the noise threshold are essential requirements for effective fault-tolerant quantum error correction protocols~\cite{sucharaQuREQuantumResource2013, bravyiHighthresholdLowoverheadFaulttolerant2024}.
One way to calibrate for high-fidelity control is to rely on an active feedback loop, also known as closed loop approach, from a real device.
This entails actively searching for actions of control that minimize error rates based on observations from the real device~\cite{baumExperimentalDeepReinforcement2021, werninghausLeakageReductionFast2021, canevaChoppedRandombasisQuantum2011, khalidSampleefficientModelbasedReinforcement2023}.
However, the cost of a real-time feedback loop with the device can be enormous.
A viable alternative approach is to develop an accurate predictive model acting as a replica of the real device.
Calibration of high-fidelity control can then be achieved with open-loop approaches, especially when combined with gradient-based algorithms for optimization in a high-dimensional search space~\cite{kochQuantumOptimalControl2022, dongQuantumControlTheory2010, anselIntroductionTheoreticalExperimental2024, omanakuttanQuditEntanglersUsing2023, khanejaOptimalControlCoupled2005, machnesTunableFlexibleEfficient2018}.

Explicitly modeling a closed-form mathematical expression of a quantum device requires comprehensive knowledge of the system as well as specialized experiments for characterization~\cite{fyrillasScalableMachineLearningassisted2024}.
The recently introduced Graybox characterization method~\cite{youssryCharacterizationControlOpen2020, youssryExperimentalGrayboxQuantum2024,youssryModelingControlReconfigurable2020,auzaQuantumControlPresence2024, mayevskyQuantumEngineeringQudits2025} addresses this issue by modeling the system as a hybrid of an explicit (Whitebox) part and an implicit (Blackbox) part.
The Whitebox part is a known (ideal) mathematical model of the system, while the blackbox part is an unknown or noisy part to be further characterized. This approach is scalable with the number of qubits, allowing us to consider only a subsystem rather than the entire system. Thus, we can apply the characterization in parallel, and we only need a single and two-qubit system at most for a set of quantum gates to be universal.

By design, the Graybox is platform-agnostic and can be applied to any physical realization of a quantum system.
However, it has so far been only used to characterize photonic platforms~\cite{youssryExperimentalGrayboxQuantum2024,youssryModelingControlReconfigurable2020}.
Extending the Graybox to other quantum computing platforms remains an open question.
Furthermore, characterization of a quantum system involves using an ensemble of measurement results to estimate expectation values of quantum observables. The estimation becomes exact only when the number of measurement samples is infinite, which is not experimentally feasible.
Therefore, one of the most significant limitations in characterization is the finite-sampling estimation of quantum observables, where the stochastic nature of quantum measurement leads to numerical uncertainty in the prediction of Graybox models.

\begin{figure*}[ht]
    \centering
    \includegraphics[width=1.0\linewidth]{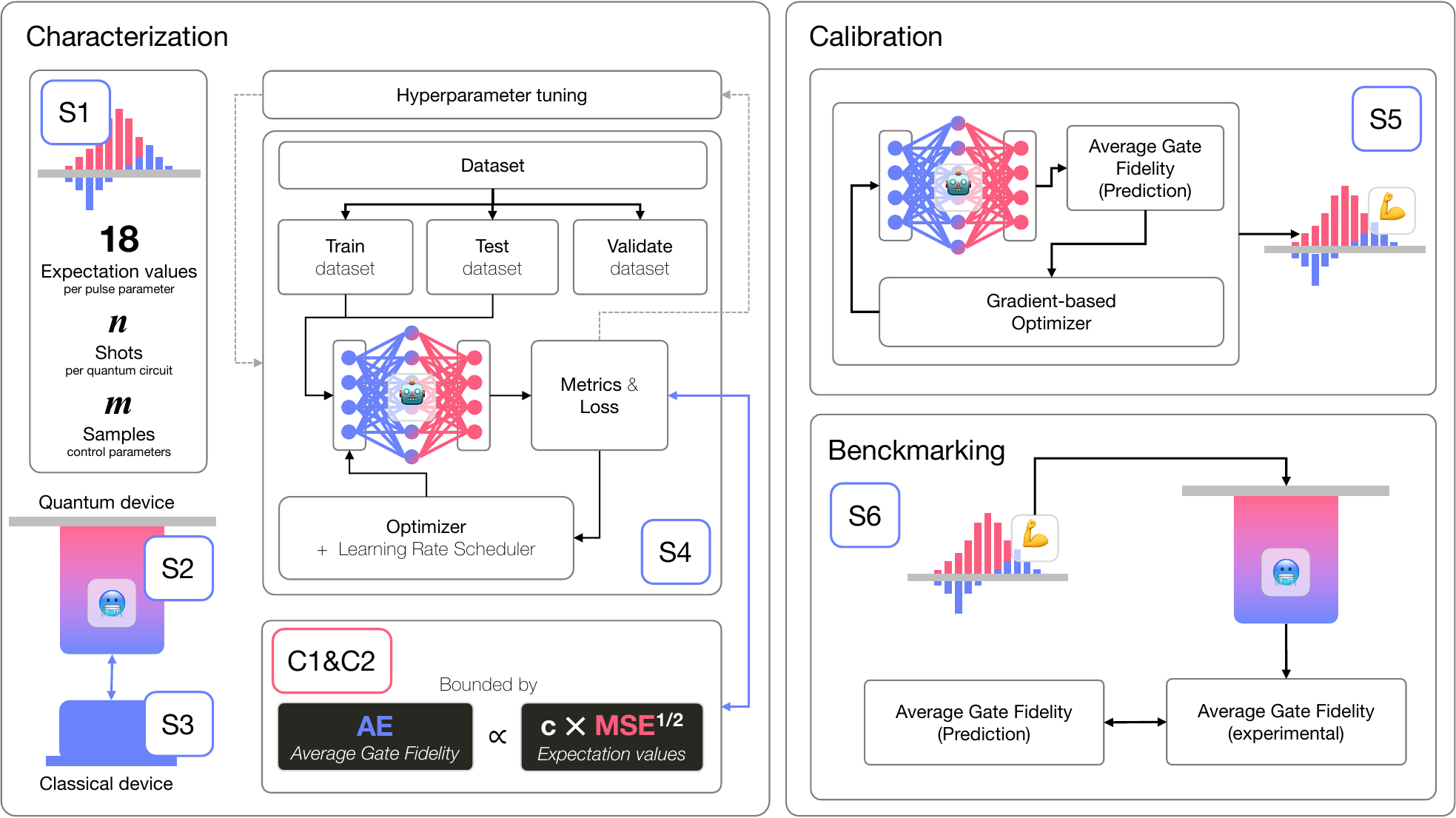}
    \caption{
        Overview of the characterization, calibration, and benchmark pipeline for a quantum device using the Graybox approach and our contribution to the approach with analysis tooling. (S1) The experiment configuration and dataset must be defined before the characterization. (S2) Data collected from the experiment or simulation. (S3) Data preprocessing which includes pre-calculation of ideal evolutions given control variables. (S4) Model training is performed with the dataset, and hyperparameters are tuned.
        (C1\&C2) We derive a bound of the expected MSE of expectation values achievable by the Graybox model, informing us when to the stop model selection. We derive bound of the expected absolute error of the average gate fidelity between the predicted and exact experimental values.
        (S5) The trained model is then used as a characterization of quantum device, i.e., it is used as a clone of the device on a local classical computer, where an open-loop control calibration is performed to acquire optimal control variables. (S6) Benchmarking of the optimized control variables is performed on the target device.
    }
    \label{fig:flow}
\end{figure*}

In this work, we address these issues by developing a framework for adapting the Graybox to the superconducting qubit platform, and experimentally characterize a superconducting qubit on \texttt{ibm\_kawasaki}, one of the IBM Quantum processors. The overview of our work is illustrated in \cref{fig:flow}
For the Blackbox part of the Graybox, implicit representation of the noise is possible with regression models such as deep neural networks (DNNs).
We observe that using the demanding long-short term memory-based NN architecture for the Blackbox as suggested in~\cite{youssryCharacterizationControlOpen2020} is not necessary in the case of the superconducting platform.
Instead, applying a simple multi-layer perceptron (MLP) is sufficient for experimental qubit characterization.
For the superconducting qubit platform, we supplement the Graybox approach by incorporating a mitigation technique addressing qubit leakage, utilizing the derivative removal by adiabatic gate (DRAG) technique \cite{motzoiSimplePulsesElimination2009} on control pulse sequences by adding the derivative of the control envelope.

We then analyze the impact of the finite-sampling estimation from the Graybox approach, which can be divided into two phases.
First, in the characterization phase, the finite-shot experimental dataset places a limit on the minimum value of loss functions achievable by the predictive model \cite{genoisQuantumOptimalControl2024,youssryMultiaxisControlQubit2023}.
Using the decomposition of MSE expected loss, we explicitly derive the irreducible error of the expected mean squared error (MSE) of expectation values between predicted values and experimental values as a function of the number of measurement shots.
Second, in the calibration phase, the average gate fidelity (AGF) becomes of great interest, especially in the context of optimal control.
We derive a bound on the Absolute Error of the AGF via the MSE bound of expectation values.
The result shows explicitly that the residue of the expected squared error of the expectation value between the exact and predicted values becomes an upper bound on the difference between the predicted and experimental AGF.
The bound also suggests that the predictive model trained using finite-sampling estimation values is sufficient to minimize the difference between predicted and experimental value to zero.
These bounds are crucial in evaluating the reliability of any predictive model trained on a finite-shot dataset so that additional information can be incorporated to improve the efficiency of the characterization and calibration experiment.

The paper is organized as follows. We first set up our mathematical foundation in \cref{sec:analytical}, starting from a general predictive model of a quantum device in \cref{sec:foundation} and deriving the expected minimum MSE of expectation values with a finite number of shots in \cref{sec:model-learnability}. We then derive the upper bound of absolute error of AGF in the form of MSE of expectation values in \cref{sec:performance-bound}. The relevant mathematical foundation of the Graybox characterization method is reviewed in \cref{sec:graybox}. Before proceeding to numerical and experimental results, we discuss our experimental procedure of predictive-model construction in \cref{sec:pipeline}. In \cref{sec:superconducting-specific}, we discuss an ideal model of the superconducting qubit device and the control of choice to be used with the Graybox characterization method. We validate the bounds from \cref{sec:model-learnability} by performing numerical simulation with ideal case conditions, where only finite-shot effects are present. We then fully characterize and calibrate the IBM superconducting device using the Graybox method and the open-loop optimization in \cref{sec:experiment}. We finally conclude our results in \cref{sec:conclusion}.

\section{Analytical Framework} \label{sec:analytical}

\subsection{Predictive model} \label{sec:foundation}

Characterization of a quantum device is a process to identify the mechanism of the device quantitatively in order to predict or control its dynamics. In real-world settings, the system's mechanism can usually be described by a mathematical model with known and unknown (or uncertain) parameters.
We then assume a parametrized model that closely imitates the mechanics of the target system and perform experiments through a form of controlled actions in order to observe the system's responses. Based on the responses, we estimate the model parameters.
We then accept any model parameters to construct a predictive model if only their predictions align with the response behavior under some acceptable error rates.

Let us construct a mathematical model to describe the dynamics of an experimental device of interest.
We assume that the device can be controlled by a set of control actions denoted by $\control = (\theta_1, \theta_2, \ldots, \theta_n)$.
In the experiments, we can then apply control actions to the device and estimate expectation values of a set of quantum observables $\hat{O}$ given initial states $\rho_0$. That is, we aim to characterize the quantum device in the form of a predictive model, $\check{f} (\control, \hat{O}, \rho_0)$, which will best estimate a true function

\begin{equation} \label{eq:predictive-model}
    f (\control, \hat{O}, \rho_0) = \langle \hat{O} \rangle _{\rho_0, \control},
\end{equation}
by minimizing some expected loss functions of interest. We assume that the predictive model can predict the expectation value of the quantum observable $\hat{O}$, given the initial state $\rho_0$ and the control variable $\control$. We can use this technique to construct predictive models that characterize quantum noisy channels. In the context of characterization for control calibration, the predictive model \cref{eq:predictive-model} can be used to calibrate elementary quantum gates at the quantum circuit level, where the objective, such as an average gate fidelity, can be calculated from predicted expectation values.

The predictive model of the form in \cref{eq:predictive-model} allows several device modeling methods to be used. The model can be constructed from explicit Hamiltonian equations with specific system parameters, or on the other hand; the model can be learned implicitly by using a statistical learning method. Fortunately, a combination of both can be used to model the device, for instance, the Graybox characterization method \cite{youssryCharacterizationControlOpen2020} that we consider in this work.

\subsection{Fundamental Limitations of Predictive Model} \label{sec:model-learnability}

Let us consider quantum experiments composed of qubits where only binary measurement outcomes are observed, e.g., $+1$ and $-1$ for qubit measurements. The expectation value given an initial state and a set of control actions, $\langle \hat{O} \rangle _{\control}$, can be estimated by taking an average of the measurement outcomes over $n$ executions of the same circuit. We refer to $n$ as the number of shots or simply shots and denote the $n$-shot estimate of the expectation value by $\mathbb{E}_n[\hat O]_{\control}$. The finite number of shots is one of the sources of uncertainty in the experimental setting, causing the model in \cref{eq:predictive-model} to be inaccurate. Typically, one would need to minimize a loss function as low as possible; however, with the finite-shot training data, there might be limits on the accuracy of the trained model and its predictive performance. In this section, we will answer the question of what one can expect in model training given a finite-shot dataset.

In order to search for an optimal predictive model for \cref{eq:predictive-model}, we consider a standard choice for the loss function in machine learning for regression tasks, which is the square error (SE). The model $\check{f}(\control) = \langle \hat{O} \rangle^{\mathrm{model}}_{\rho_0, \control}$ (noting that we have dropped the dependency on the initial state $\rho_0$ and observable $\hat{O}$ for notational simplicity)  should predict the expectation value as close as possible to the exact (infinite-shot)
expectation value $\langle \hat{O} \rangle _{\rho_0, \control}^{\rm exp}$ by learning from the experimentally obtained (finite-shot) expectation value, $\mathbb{E}_n[\hat{O}]^{\mathrm{exp}}_{\rho_0, \control}$, for a given control $\control$. For the SE loss function, we write the loss as a squared difference between the two quantities,
\begin{equation} \label{eq:msee}
    \mathcal{L}_{\mathrm{SE[E]}} \equiv \frac{1}{M} \sum_{k, \rho_0} \left( \langle \hat{O} \rangle^{\mathrm{model}}_{\rho_0, \control} -  \mathbb{E}_n[\hat{O}_{k}]^{\mathrm{exp}}_{\rho_0, \control} \right)^2,
\end{equation}
where the purpose of the sum is to average over a set of $M$ different combinations of observables $\hat O_k$ as well as initial states $\rho_0$.
We use the subscript SE[E] to indicate that the loss is an average squared error of the expectation values of the quantum observable, where the mean here refers to the average over the combinations of initial states and observables.
To approximate the lowest possible value of $\mseeloss$ given the finite-shot data, we use the decomposition
of the square error \cite{bishopPatternRecognitionMachine2006}, applying to an example of the predictive model similar to $\check{f}(\control)$ in \cref{eq:predictive-model}.
Let us consider the input $X$ as a random variable sampled from a probability distribution $X \sim p_{X}(x)$, and the output $Y$ as a random variable sampled from a probability distribution $Y \sim p_{Y|X}(y|x)$. The predictor $\check{f}(x) = \check{f}(x; \cal{D})$ (not necessary) trained on the fixed size dataset $\mathcal{D} = \{ (X, Y) \}$ and kept fixed should minimize a mean square error given by,
\begin{align} \label{eq:bias-var-decomp-alt}
    \mathbb{E}_{x, y}\!\big[ ( y - \check{f}(x))^2 \big] = \mathbb{E}_{x}\big[ (\check{f}(x) - \mathbb{E}[y|x] )^2 \big] +  \mathbb{E}_{x}\big[ \mathrm{Var}(y|x) \big].
\end{align}
The expectations in \cref{eq:bias-var-decomp-alt} are with respect to all possible values of input and output. Note that both terms are positive.
\cref{eq:bias-var-decomp-alt} is minimized when the predictor is optimal, in other words it predicts the expected value of the output for a given input $\check{f}(x) = \mathbb{E}[y|x]$, reducing the first to zero.
However, the last term, the expectation of conditional variance of the distribution of $Y$ given the input value of $X = x$ over input space $x$, is considered an irreducible error as it is inherited in the observed data $Y$. Thus, the expected conditional variance is the lower bound of the square error, $\mathbb{E}_{x,y}\!\big[ ( y - \check{f}(x))^2 \big]$, achievable by the optimal predictor.

In our context of \cref{eq:predictive-model},
our model $\check{f}(\control)$ should minimize $\mseeloss$ in \cref{eq:msee} where the observed data is the finite-shot expectation value, i.e., $Y = \mathbb{E}_n[\hat O]_{\rho_0,\control}^{\rm exp}$.
Therefore, we can apply the decomposition in \cref{eq:bias-var-decomp-alt}, replacing $x$ with $\control$, $\check{f}(x)$ with $\langle \hat{O} \rangle^{\mathrm{model}}_{\rho_0, \control}$, taking the average over all possible initial states and observables. We find that the conditional variance in \cref{eq:bias-var-decomp-alt} is given by
\begin{equation}\label{eq:msee-boundby-var}
    \mathrm{Var} (y|x) \mapsto \mathrm{Var} \left(\mathbb{E}_n[\hat O]_{\rho_0,\control}^{\rm exp} | \control\right),
\end{equation}
which, the expectation over the control variable should give us an approximated lower bound for the loss in \cref{eq:msee}. The conditional variance $\mathrm{Var} (y|x)$ could be extracted directly from the experimental data. Given particular control parameters, observables, and initial states, a large number of experiments need to be executed so that one can approximate the exact conditional variance. Then, the sample variance should be averaged over all combinations of initial states and observables. However, this method is costly as it requires additional experiments just to estimate the conditional variance. We instead derive an analytical expression of the conditional variance using the variance of a quantum observable based on Bernoulli distribution.
The variance of an estimation of an observable $\hat{O}$ from $n$ measurement shots is
\begin{equation} \label{eq:var-observable}
    \mathrm{Var}\left(\mathbb{E}_n[\hat O]_{\rho_0,\control}^{\rm exp} | \control \right) = \frac{1}{n} (1 - \langle\hat{O}\rangle_{\rho_0,\control}^2 ),
\end{equation}
given the control variables $\control$ and the initial state $\rho_{0}$ fixed when evaluating the expectation value.

We then calculate the bound for our single qubit case, where, we consider a set of observables $\hat{O} \in \{ \hat{X}, \hat{Y}, \hat{Z}\}$ which fully characterize a quantum state and a set of pure initial states $\rho_0 \in \{ |+\rangle \langle +|, |-\rangle \langle -|, |i\rangle \langle i|, |-i\rangle \langle -i|, |0\rangle \langle 0|, |1\rangle \langle 1| \}$. Then, assuming that the control $\control$ only produces unitary operations and all initial states are pure states, we can apply $\langle X \rangle^2_{\rho_{0}, \control} + \langle Y \rangle^2_{\rho_{0}, \control} + \langle Z \rangle^2_{\rho_{0}, \control} = 1$ and obtain the lower bound of an expected loss of \cref{eq:msee} as
\begin{align}\label{eq:var-data}
    \mathbb{E}_{x,y}\!\left[{\cal L}_{\rm SE[E]} \right] & \ge \mathbb{E}_{x}\!\left[
    \frac{1}{18n} \sum_{\rho_{0}} \sum_{k} ( 1 - \langle\hat{O}_k\rangle_{\rho_{0}, \control}^2 ) \right] \nonumber                         \\
                                                           & = \mathbb{E}_{x}\!\left[ \frac{1}{18n} \sum_{\rho_0} 2 \right] = \frac{2}{3n}.
\end{align}
We refer to the average of the conditional variances (irreducible errors), which is independent of the choice of control action as \emph{data variance}. We note that the result does not depend on the number of initial states $\rho_0$ taken in the averaging process.

We now discuss how to use the data variance in practice.
Since, the expected loss on the LHS of \cref{eq:bias-var-decomp-alt} is the expectation with respect to joint distribution $p_{X, Y}(x, y) = p_{Y|X}(y|x)p_{X}(x)$, we can estimate it by averaging the squared error over samples from the distribution $p_{X, Y}(x, y)$. Since the $x,y$ samples will form a dataset, we denote $\mathbb{E}_{\{X, Y \}}[\cdot] \rightarrow \mathbb{E}_{\rm dataset}[\cdot]$.
In the subsequent calculations, the samples can be taken from the training or testing dataset.
Then, we can calculate the LHS of \cref{eq:bias-var-decomp-alt} by averaging the squared error over the dataset using the predictive model.
Consequently, the optimality of the predictor, i.e., first term on the RHS of \cref{eq:bias-var-decomp-alt}, can be calculated from the empirical value of expected loss, i.e. LHS of \cref{eq:bias-var-decomp-alt}, subtracted by the data variance.
Thus, the data variance, which is an analytical expression of the irreducible error, would provide additional information, complementing the empirical result in the analysis of the predictive model.
For instance, we could design an optimization stopping criterion in the predictive model construction process based on the empirical optimality of the predictor. 

With \cref{eq:var-data}, we know that the expected loss should be at most reduced to $2/3n$, which also decreases as the number of shots $n$ increases. In practice, we construct a predictive model by minimizing \cref{eq:bias-var-decomp-alt} over the training dataset $\mathbb{E}_{\rm train}[\mseeloss]$. However, care must be taken when selecting a model based on expected training loss. For example, in the case of using a neural network as a part of a predictive model, it may be possible that the model is large enough to overfit the training dataset, but the expected loss approach $2/3n$ results in selecting a non-optimal model. Thus, it is considered a good practice to evaluate the expected loss using (unseen) testing dataset $\mathbb{E}_{\rm test}[\mseeloss]$. Therefore, we expect that selecting a model that predicts the expected loss over the training and testing datasets simultaneously, which also approximately approaches the data variance, is a first guiding step in model selection. 

\subsection{Performance Bounds for Control Optimization} \label{sec:performance-bound}

From the previous section, the predictive model is trained to predict the expectation value as closely as possible to the experimental data $\mathbb{E}_{\mathrm{train}}[\mseeloss] \approx \mathbb{E}_{\mathrm{test}}[\mseeloss]$ and also approach the data variance.
However, the MSE of the expectation value does not provide a straightforward and meaningful interpretation of model performance.
For the control calibration problem, we are interested in how close an average gate fidelity (AGF) predicted by the model is to the exact value from the experiment, showing that the predictive model can be used as a local replica of the device.
This is in contrast to~\cite{khalidSampleefficientModelbasedReinforcement2023}, where authors studied the sample efficiency as a function of Hamiltonian error.

To make the connection between AGF and the predictive model clear, we will rewrite the AGF as a function of expectation values. Let us start with the definition of AGF between an ideal unitary $\hat{U}$ parametrized by control variable $\control$ (not shown explicitly) and a map $\mathcal{E}$. Let $\rho = |\psi\rangle \langle\psi|$ be an arbitrary pure state,
an average of state fidelity between $\hat{U}^{\dagger} | \psi \rangle \langle \psi | \hat{U}$ and $\mathcal{E} (| \psi \rangle \langle \psi |)$ integrating over all possible $|\psi\rangle$, which is
\begin{equation} \label{eq:agf-def}
    \bar{F}(\mathcal{E}, \hat{U}) = \int d\psi \langle \psi | \hat{U}^{\dagger} \mathcal{E} (| \psi \rangle \langle \psi |) \hat{U} | \psi \rangle.
\end{equation}
Typically, we are interested in the case where the map $\mathcal{E}$ is an experimental realization of $\hat{U}$. \blk
For the single qubit case, the system dimension is $d = 2$, and the AGF can be written \cite{nielsenSimpleFormulaAverage2002} with a state decomposition on the Pauli matrices as
\begin{subequations}
    \label{eq:agf}
    \begin{align}
        \bar{F}(\mathcal{E}, \hat{U}) = & \frac{1}{2} + \frac{1}{12} \sum^3_{j=1} \sum^1_{k=0} \alpha_{j,k} \tr{ \left[ \hat{U} P^{\dagger}_j \hat{U}^{\dagger} \mathcal{E} ( \rho_{j,k} ) \right] } \label{eq:direct-agf} \\
        =                               & \frac{1}{2} + \frac{1}{12} \sum^3_{j,m=1} \sum^1_{k=0} \alpha_{j,k} \beta_{m, j} \tr{ \left[ \hat{P}_m \mathcal{E} ( \rho_{j,k} ) \right] }, \label{eq:agf-experiment}           \\
        =                               & \frac{1}{2} + \frac{1}{12} \sum^3_{j,m=1} \sum^1_{k=0} \alpha_{j,k} \beta_{m, j} \langle \hat{O} \rangle_{j, k, m} \label{eq:direct-agf-2}
    \end{align}
\end{subequations}
where indices $j, m$ indicate different Pauli observables, i.e., $\hat{P}_j, \hat{P}_m \in \{\hat{X}, \hat{Y}, \hat{Z}\}$, and $\rho_{j,k} \in \{ |+\rangle \langle +|, |-\rangle \langle -|, |i\rangle \langle i|, |-i\rangle \langle -i|, |0\rangle \langle 0|, |1\rangle \langle 1| \}$.
In this notation, $j$ indicates the Pauli observable and $k$ indicates whether the state is a +1 or a -1 eigenstate of that observable.
For example, $\rho_{2,1}=|-i\rangle\langle-i|$.
For any Pauli observable $\hat{P}_j$, we have $\alpha_{j,0} = 1$ and $\alpha_{j,1} = -1$.
From \cref{eq:direct-agf}, we substitute $\hat{U} P^{\dagger}_j \hat{U}^{\dagger}$ with its Pauli basis expansion, resulting in \cref{eq:agf-experiment} where $\beta_{m, j} = \frac{1}{2} \tr{\left[ \hat{U} \hat{P}^{\dagger}_j \hat{U}^{\dagger} \hat{P}_{m} \right] }$. This expansion allows us to estimate the AGF between any ideal (qubit) unitary map and its experimental realization by measuring the expectation values of at most 18 combinations of 6 initial states and 3 Pauli observables.
Notice that $\tr{ \left[ \hat{P}_m \mathcal{E} ( \rho_{j,k} ) \right] }$ is the exact experimental expectation value, so we denote it as $\langle \hat{O} \rangle_{j, k, m}$ in \cref{eq:direct-agf-2}.

Since the predictive model $\check{f}(\control)$ can predict the expectation values of a given control parameter, one can calculate the AGF from the prediction. The quantity of interest is the accuracy of AGF prediction, $\bar{F}_{\mathrm{model}}$, to the exact value of the experimental realization, $\bar{F}_{\mathrm{exact}}$. Here, we note that $\bar{F}_{\mathrm{exact}}$ is not a finite-shot estimation of AGF measurable in experimental realization, but is the infinite-shot (exact) value.
The difference between them is obtained using \cref{eq:agf-experiment},
\begin{align} \label{eq:agf-diff}
    \bar{F}_{\mathrm{model}} - &\bar{F}_{\mathrm{exact}} =\nonumber\\
    &\frac{1}{12} \sum^3_{j,m=1} \sum^1_{k=0} \alpha_{j,k} \beta_{m, j} ( \langle \hat{O} \rangle_{j, k, m}^{\rm model} - \langle \hat{O} \rangle_{j, k, m}^{\mathrm{exact}} ),
\end{align}

We now show that the AGF difference in \cref{eq:agf-diff} and the MSE in \cref{eq:msee} are related through the Cauchy-Schwarz inequality,
\begin{equation} \label{eq:cauchy-schwarz}
    \left( \sum_i u_i v_i \right)^2 \leq \left( \sum_i u_i^2 \right) \left( \sum_i v_i^2 \right).
\end{equation}
Let us start by defining functions
\begin{subequations}
    \label{eq:uivi}
    \begin{align}
        u_{i} = \,& \langle \hat{O} \rangle_{i}^{\rm model} - \langle \hat{O} \rangle_{i}^{\mathrm{exact}} \equiv \langle \hat{O} \rangle_{j, k, m}^{\rm model} - \langle \hat{O} \rangle_{j, k, m}^{\mathrm{exact}}, \label{eq:ui} \\
        v_{i} \equiv \,& \alpha_{j,k} \beta_{m, j} \label{eq:vi},
    \end{align}
\end{subequations}
where the indices $j,k,m$ are combined into single index $i$,
which leads to writing \cref{eq:agf-diff} as
\begin{equation}\label{eq:lhs-bound}
    \bar{F}_{\mathrm{model}} - \bar{F}_{\mathrm{exact}} = \frac{1}{12} \sum_i v_i u_i.
\end{equation}
Note that we use the Cauchy-Schwarz inequality for real numbers, so we have to show that $u_i$ and $v_i$ are real numbers. Since $u_i$ is defined in terms of the expectation value of an observable, it must be a real number. For $v_i$, we already know that $\alpha_{j, k} \in \{-1, 1\}$, so we have to show that $\beta_{m, j}$ is a real number. We consider $\beta_{m,j}^{*}$ as follows,
\begin{subequations}
    \begin{align}
        \left(\frac{1}{2} \tr{\left[ \hat{U} \hat{P}^{\dagger}_j \hat{U}^{\dagger} \hat{P}_{m} \right]} \right)^{*}
                        & = \frac{1}{2} \tr{\left[ (\hat{U} \hat{P}^{\dagger}_j \hat{U}^{\dagger} \hat{P}_{m})^{\dagger} \right] },      \\
                        & = \frac{1}{2} \tr{\left[ \hat{U} \hat{P}^{\dagger}_j \hat{U}^{\dagger} \hat{P}_{m} \right]}, \label{eq:beta-3} \\
        \beta_{m,j}^{*} & = \beta_{m,j} \label{eq:beta-4}
    \end{align}
\end{subequations}
where, we use the cyclic property of trace, and use the fact that $\hat{P}_m = \hat{P}_m^{\dagger}$ and $\hat{P}_j = \hat{P}_j^{\dagger}$ in \cref{eq:beta-3}. From \cref{eq:beta-4}, we show that $\beta_{m, j}$ must be a real number. Therefore, $v_{i}$ must be a real number.
Now, we will express \cref{eq:msee} in terms of $u_i$. We denote the experimental data $ \mathbb{E}_n[\hat{O}_{k}]^{\mathrm{exp}}_{\rho_0, \control}$ with $\mathbb{E}_{i}^{\mathrm{exp}}$ for notation brevity,
\begin{align}
    \mathcal{L}_{\mathrm{SE[E]}} & = \frac{1}{18} \sum_{i} \left( \langle \hat{O} \rangle_{i}^{\rm model}  - \mathbb{E}_{i}^{\mathrm{exp}} \right)^2.
\end{align} \label{eq:msee-loss-i}
We then express \cref{eq:msee-loss-i} in terms of $u_i$ as follows,
\begin{subequations}
    \label{eq:sum-ui}
    \begin{align}
        \mathcal{L}_{\mathrm{SE[E]}} & = \frac{1}{18} \sum_{i} \left( \langle \hat{O} \rangle_{i}^{\rm model} - \langle \hat{O} \rangle_{i}^{\mathrm{exact}} + \langle \hat{O} \rangle_{i}^{\mathrm{exact}}  - \mathbb{E}_{i}^{\mathrm{exp}} \right)^2 \label{eq:se-expand-1} \\
                                     & = \frac{1}{18} \sum_i (u_i - \varepsilon_{i})^2 \label{eq:se-expand-2}                                                                                                                                                                 \\
                                     & = \frac{1}{18} \sum_i u_i^2 - \frac{1}{18} \sum_i 2 u_i \varepsilon_i + \frac{1}{18} \sum_i \varepsilon_i^2. \label{eq:se-expand-3}
    \end{align}
\end{subequations}
In \cref{eq:se-expand-1}, we inserted zero in the forms of the exact experimental expectation value $\langle \hat{O} \rangle_{i}^{\mathrm{exact}}$. Then, in \cref{eq:se-expand-2}, we defined $\varepsilon_i = \mathbb{E}_{i}^{\mathrm{exp}} - \langle \hat{O} \rangle_{i}^{\mathrm{exact}}$ and the remaining terms are $u_i$ as we defined in \cref{eq:ui}. Lastly, we expanded the expression in \cref{eq:se-expand-3}. We then, rearrange for $\sum_i u_i^2$ as follows,
\begin{align}
    \sum_i u_i^2 = 18 \mathcal{L}_{\mathrm{SE[E]}} + \sum_i (2 u_i \varepsilon_i - \varepsilon_i^2).
\end{align}

For the summation $\sum_{i} v_{i}^{2}$, we can look at the properties of $\alpha_{j,k}$ and $\beta_{m, j}$. Since we know that $\alpha_{j, k}$ has two possible values, $\alpha_{j,0} = 1$ and $\alpha_{j,1} = -1$, and that $\beta_{m, j}$ does not depend on the index $k$, we can write
\begin{equation}
    \sum_i v_i^2 = \sum_{j, k, m} (\alpha_{j,k} \beta_{m, j})^2 = 2 \sum_{j, m} (\beta_{m, j})^2.
\end{equation}
This quantity can be simplified even further. Note that $\beta_{m, j} = \frac{1}{2} \tr{\left[ \hat{U} \hat{P}^{\dagger}_j \hat{U}^{\dagger} \hat{P}_{m} \right]  }$, where $\hat{P}_j$ and $\hat{P}_m$ are Pauli matrices.
We express $\hat{P}_j^\dagger = \hat{P}_j = \rho_{j,+} - \rho_{j, -}$ in terms of its $+1$ and $-1$ eigenvectors.
When the initial states $\rho_{+}$ and $\rho_{-}$ evolve under the same unitary, their expectation values when measuring the same Pauli observable will have the same magnitude but opposite sign, $\tr{ \left[ \hat{U} \rho_{j, +} \hat{U}^\dagger \hat{P}_m \right] } = - \tr{ \left[  \hat{U} \rho_{j, -} \hat{U}^\dagger \hat{P}_m \right] }$
We can therefore write
\begin{align} \label{eq:bmj-1}
    \beta_{m, j} & =  \tr{ \left[  \hat{U} \rho_{j, +} \hat{U}^\dagger \hat{P}_m \right] } = \langle \hat{P}_m \rangle_{\rho_{j, +}}.
\end{align}
Note that the quantity we want is the summation $\sum_{j, m} (\beta_{m, j})^{2}$, which, given \cref{eq:bmj-1}, we see that the summation with the index $m$ leads to the exact form of purity of a state $\hat{U} \rho_{j, +} \hat{U}^{\dagger}$.
Given that we know $\rho_{j, +}$ is a (pure) eigenstate, we are sure that the purity is unity. That is, the summation of index $m$ will be $1$. Since index $j$ is the sum over initial states, this means that the summation with the index $j$ gives a multiplicative factor of $3$, leading to
\begin{equation} \label{eq:sum-vi}
    \sum_i v_i^2 = 2 \sum_{j, m} (\beta_{m, j})^2 = 2\sum_{j, m} \langle \hat{P}_m \rangle_{\rho_{j, +}}^2 = 6.
\end{equation}
Substituting the results of $u_{i}$ and $v_{i}$ summation into \cref{eq:cauchy-schwarz}, we arrive at the  following inequality,
\begin{equation} \label{eq:performance-bound}
    (\bar{F}_{\mathrm{model}} - \bar{F}_{\mathrm{exact}}) ^ 2 \leq \frac{6}{12^2} \left(18 \mseeloss + \sum_i (2 u_i \varepsilon_i - \varepsilon_i^2)\right).
\end{equation}
To remove the random variable $\varepsilon_i$, we compute the expected value of the inequality. Notice that, $u_i$ and $v_i$ do not depend on $\varepsilon$, so $\bar{F}_{\mathrm{model}} - \bar{F}_{\mathrm{exact}}$ is constant with respect to $\mathbb{E}_{y|x}$. We use the fact that $\mathbb{E}_{\varepsilon}[\varepsilon_i] = 0$, and notice that $\mathbb{E}_{y|x}[\varepsilon_i^2] = \mathrm{Var} \left(\mathbb{E}_n[\hat O]_{\rho_0,\control}^{\rm exp} | \control\right)$ is the conditional variance with the result from \cref{eq:var-data} to obtain the following bound,
\begin{align} \label{eq:before}
    (\bar{F}_{\mathrm{model}} - \bar{F}_{\mathrm{exact}}) ^ 2 \leq \frac{3}{4} \left(\mathbb{E}_{y|x}[\mseeloss] - \frac{2}{3n}\right).
\end{align}

Furthermore, we can connect \cref{eq:bias-var-decomp-alt} to the absolute difference between the AGF predicted by model and exact experimental value by considering an expected value of inequality in \cref{eq:before} with respect to $x$ and using the Jensen inequality for a random variable $|\bar{F}_{\mathrm{model}} - \bar{F}_{\mathrm{exact}}|$ with respect to $x$. We then have
\begin{align} \label{eq:agf-bound-alt}
    \mathbb{E}_{x} \big[ |\bar{F}_{\mathrm{model}} - \bar{F}_{\mathrm{exact}}| \big] \leq \sqrt{ \frac{3}{4} \left(\mathbb{E}_{x, y}[\mseeloss] - \frac{2}{3n} \right)},
\end{align}
where the inequality becomes the bound of the expected difference of AGF over different inputs in terms of the expected loss.

With decomposition of $\mseeloss$, the upper bound of the AGF difference above depends only on the $\mathbb{E}_{x}\big[ (\check{f}(x) - \mathbb{E}[y|x] )^2 \big]$ since the dependence on $n$ is cancelled out. Note that we can estimate $\mathbb{E}_{x}\big[ (\check{f}(x) - \mathbb{E}[y|x] )^2 \big]$ using the testing dataset discussed in previous section.
This result suggests that a high-precision quantum gate can be reliably produced by using a predictive model trained on a few-shot dataset.
In the best case scenario, with optimal predictive model $\mathbb{E}_{x,y}[\mseeloss] = \frac{2}{3n}$, the upper bound on the RHS becomes zero, so the model can predict the exact experimental AGF.
We would like to highlight that the LHS of \cref{eq:agf-bound-alt} is the difference between the model prediction and the exact experimental value, and is not the difference between the model prediction and the finite-shot experimental value.
These analytically derived bounds in \cref{eq:var-data}, \cref{eq:before}, and \cref{eq:agf-bound-alt} confirm the observed behaviour in \cite{youssryMultiaxisControlQubit2023}.

This result leads us to the following observation: if the application of interest is to use the predictive model to predict AGF, then instead of trying to optimize for the upper bound of AGF difference, we can directly optimize for the absolute tolerance between experimental AGF $\bar{F}_{\mathrm{exp}}$ and predicted AGF $\bar{F}_{\mathrm{model}}$, i.e., an absolute error (AE) loss function of predicted and finite-shot experimental AGF
\begin{align} \label{eq:maef}
    \mathcal{L}_{\mathrm{AE[F]}} = | \bar{F}_{\mathrm{model}} - \bar{F}_{\mathrm{exp}}|.
\end{align}
We will compare model training using $\mseeloss$ and $\maefloss$ in \cref{sec:experiment} and discuss the trade-off between the two.

\subsection{Graybox: A Predictive Model for Quantum Device Characterization and Control} \label{sec:graybox}

To demonstrate the utilization of the analytical results built from the predictive model in \cref{eq:predictive-model} presented in \cref{sec:model-learnability} and \cref{sec:performance-bound}, we consider the Graybox characterization method. The Graybox model comprises (1) Whitebox, the explicit ideal Hamiltonian of the target device, and (2) Blackbox, the implicit form of noise represented by the machine learning model. The Graybox characterization method has an advantage that it is independent of the physical realization of quantum devices and it is flexible in the noise model, while the prediction respects quantum dynamics imposed by Whitebox. Even though, the Graybox approach can be extended to multiple-qubit cases, in the context of control calibration, only single-qubit and two-qubit quantum gates are required for a universal gateset. Thus, we expect that the Graybox method does not suffer from the exponential growth requirement of memory and computation time for larger qubit systems. In our case, we consider a single qubit system as a simple example.
In this section, we will detail the mathematical model relevant to the Graybox approach following \cite{youssryCharacterizationControlOpen2020} and discuss its applicability to the characterization of quantum devices.

The Graybox approach assumes that the device's Hamiltonian can generally be decomposed into the parametrized ideal and noisy parts,
\begin{equation}\label{eq:total-hamiltonian}
    \hat{H}_{\mathrm{total}} (\control, t) = \hat{H}_{\mathrm{ideal}} (\control, t) + \hat{H}_{\mathrm{noise}} (t),
\end{equation}
where $\hat{H}_{\mathrm{ideal}}$ is the device Hamiltonian constructed from the ideal device parameters. For example, in a superconducting qubit experiment, the ideal parameters are the qubit's frequencies and the input control variable $\control$, assuming no effect from noise.

Given this general form of the Hamiltonian, one can see that the effect of noise in $\hat{H}_{\mathrm{noise}}(t)$, could be removed or canceled by choosing a proper controlled Hamiltonian, $\hat{H}_{\mathrm{ideal}} (\control, t)$. We will show later that, with the Graybox approach, no additional assumptions for the noise, such as the forms of the noise Hamiltonian, are needed.

We write the quantity that can be observed in experiments as the expectation value of a system's observable $\hat{O}$ at time $T$,
\begin{equation} \label{eq:general-expval}
    \mathbb{E} [\hat{O}(T)]_{\rho_{0}, \control} \equiv \left\langle \mathrm{Tr} \left[ \hat{U}(\control, T) (\rho_{0} \otimes \rho_{B}) \hat{U}^{\dagger} (\control, T) \hat{O} \right] \right\rangle _{c},
\end{equation}
where the total unitary operator is $\hat{U}(\control, T) = \mathcal{T}_+ \mathrm{exp} \left\{ -i \int_{0}^T \hat{H}_{\mathrm{total}}(\control, s) ds \right\}$ as a time-ordered propagator of the total Hamiltonian, and we have assumed unentangled initial state of the system and its environment (bath), $\rho_{0} \otimes \rho_{B}$. The trace is over both the system-bath Hilbert space, while $\langle\cdot\rangle_c$ is a stochastic average over possible realizations of any random variables in the total Hamiltonian, e.g., those in the noisy part $\hat{H}_{\mathrm{noise}}(t)$. We emphasize again that we do not need a particular form of the noise Hamiltonian. However, one might need to assume it for numerical simulation, e.g., in \cite{youssryCharacterizationControlOpen2020} where the noise Hamiltonian is of the form $\hat{H}_{\mathrm{noise}} (t) = \sum_{\alpha = X, Y, Z} \hat{\sigma}_{\alpha} \otimes \hat{B}_{\alpha} (t)$, where $\hat{B}_\alpha(t)$ are operators capturing the effect of the quantum bath and the classical noise, where the latter can be assumed to be colored noise generated from a Power Spectral Density (PSD).

The advantage of the Graybox characterization method lies in its formulation that separates ideal and noisy evolution. In the following, we demonstrate how the Graybox method achieves system-environment decomposition.
The technique is similar to the usual  interaction frame, but with a slight variation. Let us define the interaction Hamiltonian as
\begin{equation}\label{eq:totalHam}
    \hat{H}_{I}(t) = \hat{U}^{\dagger}_\mathrm{ideal}(t) \hat{H}_{\mathrm{noise}} (t) \hat{U}_\mathrm{ideal}(t),
\end{equation}
where the ideal unitary operator is defined as usual as $\hat{U}_\mathrm{ideal}(t) = \mathcal{T}_{+} \mathrm{exp} \{ -i \int_0^t \hat{H}_\mathrm{ideal}(s) ds \}$. The usual interaction frame would lead to a decomposition of the total unitary operator as $\hat U(\control,t) = \hat U_{\rm ideal}(t) \hat U_I(t)$, where $\hat U_I$ is the unitary operator in the interaction frame.
However, in the Graybox approach, it requires a modified interaction frame such that the decomposition becomes
\begin{equation} \label{eq:total-unitary}
    \hat{U}(\control,t) = \hat{U}_{J}(t) \hat{U}_{\mathrm{ideal}}(t).
\end{equation}
In order to achieve this form, we find that the noisy propagator $\hat U_J$ is given by~\cite{youssryCharacterizationControlOpen2020, chalermpusitarakFrameBasedFilterFunctionFormalism2021}
\begin{equation} \label{eq:noisy-propagator}
    \hat{U}_{J} (t) = \mathcal{T}_+ \mathrm{exp} \left\{ -i\int_{0}^t \hat{U}_\mathrm{ideal}(t) \hat{H}_{I}(s) \hat{U}^{\dagger}_\mathrm{ideal}(t) d s \right\},
\end{equation}
where $\hat H_I$ and $\hat U_{\rm ideal}$ are given above. Using the decomposition Eq.~\eqref{eq:total-unitary}, we substitute it back in \cref{eq:general-expval} and cyclic the operators to get
\begin{align} \label{eq:exp-wo}
    \mathbb{E} [\hat{O}(T)]_{\rho_{0}} & =
    \mathrm{Tr} \left[ \left\langle \hat{U}^{\dagger}_{J} (T) \hat{O} \hat{U}_{J} (T)  \right\rangle_c \hat{U}_{\mathrm{ideal}} (T)\rho_{0} \hat{U}^{\dagger}_{\mathrm{ideal}} (T) \right], \nonumber \\
                                       & = \mathrm{Tr} \left[ \hat{W}_{O} (T) \hat{U}_{\mathrm{ideal}} (T)\rho_{0} \hat{U}^{\dagger}_{\mathrm{ideal}} (T) \right],
\end{align}
which interestingly is in a form that the observable expectation value, $\mathbb{E} [\hat{O}(T)]_{\rho_{0}}$, is a simple trace of the ``ideal" state dynamics, $\hat{U}_{\mathrm{ideal}} (T)\rho_{0} \hat{U}^{\dagger}_{\mathrm{ideal}} (T)$, with a noise-affected observable operator defined as
\begin{align} \label{eq:wo}
    \hat{W}_{O} (T) = \left\langle \hat{U}^{\dagger}_{J} (T) \hat{O} \hat{U}_{J} (T)  \right\rangle_c.
\end{align}

The noise operator in \cref{eq:wo} depends on control because $\hat{U}_{J}$ also depends on $\hat{U}_{\mathrm{ideal}}$. Thus, we can interpret the noise operator $\hat{W}_{O}$ as a control-induced noise.
Based on the expectation value in \cref{eq:exp-wo}, we see that it can be separated into what is known to the experimenters through the controls in $\hat{U}_{\mathrm{ideal}}$, and what is unknown noise effects in \cref{eq:wo}.
The former is a Whitebox, part of our algorithm, while the rest (unknown) is referred to as Blackbox. Combining them will result in the expectation value defined in \cref{eq:exp-wo}, which results in a predictive model as defined in \cref{eq:predictive-model}. In \cite{youssryCharacterizationControlOpen2020}, they defined this combination as a Graybox.
Note that, in the ideal noiseless case, $\hat{W}_{O} = \hat{O}$. Knowing $\hat{W}_{O}$ has several benefits. First, one can evaluate how close $\hat{W}_{O}$ is to the ideal observable $\hat{O}$ using unitary fidelity and determine the physical insight of how control and system bias affect specific Pauli axes. Second, one can use the metric incorporated in model training or as a cost function for control calibration.

Next, we will discuss an implementation of the Graybox predictive model.
\cref{eq:wo} and the definitions of $\hat{U}_{I, R} (T)$ in \cref{eq:noisy-propagator} indicate that $\hat{W}_{O}= \mathcal{B}(\hat{H}_\mathrm{noise}, \mathbf{\Theta})$ is a function of the noise Hamiltonian and a vector of constrained control parameters. That is, different controls induce different noise effects. Thus, it is possible to use deep neural network, i.e., Blackbox, to learn the mapping from $\mathbf{\Theta}$ to $\hat{W}_{O}$, as it is a universal function approximator. We then have to calculate the ideal propagator, $\hat{U}_{\mathrm{ideal}}$, given control parameters, i.e., Whitebox. Calculating the perfect propagator $\hat{U}_\mathrm{ideal}$ requires the calculation of the time-ordered exponential of the device Hamiltonian, which can be done with several numerical techniques. One such technique is the first-order Trotterization method used in~\cite{youssryCharacterizationControlOpen2020}. This is a first-order approximation to the real propagator, and the approximation error is regarded as noise to be taken into account by Blackbox. Another technique is to solve the Schrödinger equation using an ODE solver.

To summarize, Blackbox predicts $\hat{W}_{O}$, and Whitebox calculates the ideal propagator given $\control$, combining both resulting in the predictive model, Graybox, i.e., characterized targeting device.

\section{Practical Implementation} \label{sec:cc}

In this section, we explain and discuss a detailed device characterization and control calibration procedure. We start by discussing the experimental procedure, independent of quantum device specification in \cref{sec:pipeline}. Next, in \cref{sec:superconducting-specific}, we provided a specification of a superconducting qubit as an example for our implementation, including the device Hamiltonian and choice of control, i.e., pulse envelope and the bounds of each parameter.
We then show numerically in \cref{sec:idealshots} that the model can be easily trained to the optimal point in the ideal case, where regression is simple and the optimal point is indeed in agreement with the analytical result in \cref{sec:model-learnability}.
Next, we characterize a real superconducting quantum computer, calibrate control variables for the \sx gate, and benchmark the results compared to our model prediction in \cref{sec:experiment}.

\subsection{Experimental procedure} \label{sec:pipeline}

\begin{figure}
    \centering
    \includegraphics[width=1.0\linewidth]{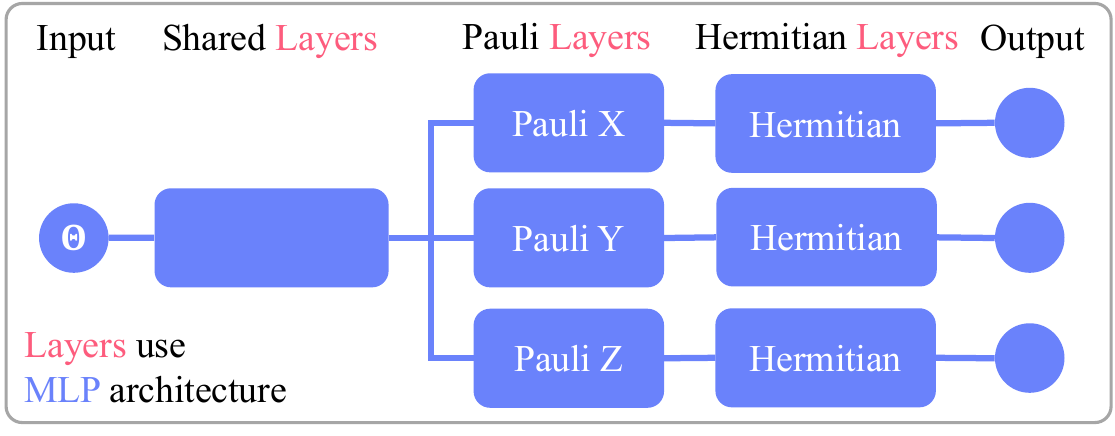}
    \caption{
        The architecture of Blackbox, a part of the Graybox predictive model. The model consists of three group of layers: Shared, Pauli, and Hermitian layers. They use MLP architecture. The control variables $\control$ are the input parameters. The Shared layer transforms the vector of input parameters into an intermediate vector. Each of the three Pauli layers is then transforming the output of the Shared layer independently. Finally, the Hermitian layer is responsible for transforming the output of the Pauli layer into parameters of the Hermitian matrix corresponding to \cref{eq:wo} for each Pauli observable.
    }
    \label{fig:model-architecture}
\end{figure}

Before proceeding to our numerical and experimental results, we outline our characterization procedure, which includes data acquisition, preprocessing, model training, and hyperparameter tuning.
We show the high-level illustration of the experimental procedure in \cref{fig:flow}.

\textbf{(S1) designing configuration phase} consists of (\textbf{R1 - R4}) requirements to be identified before performing an experiment.
\begin{enumerate}[start=1,label={(\bfseries R\arabic*)}]
    \item The specifications of qubit(s) in the quantum device are needed for composing the Whitebox, while Blackbox will characterize the noise. Qudit extension is also possible. For example, for the superconducting platform, the energy splitting of the system can be specified in the Whitebox.
    \item Functional form of control (or the pulse sequence in the superconducting qubit example) and bounds of parameters are needed. As discussed previously, these specifications depend on the device of interest.
    \item Number of samples, $m$, is the size of the dataset used to train the model. Each sample consists of the control variables (input or feature in ML literature) and 18 combinations of expectation values (output or label in ML literature) for a single qubit case. The combination comes from 6 initial states corresponding to eigenvectors of three Pauli observables, $|0\rangle, |1\rangle, |+\rangle, |-\rangle, |i\rangle, |-i\rangle$, and 3 Pauli observables.
    \item Number of shots, $n$, that define model training and prediction accuracy limitations.
\end{enumerate}

Once the experiment configuration is defined, the next step is \textbf{(S2) collection of data}. At this stage, one can collect data by experimenting on a real device or a simulation with or without noise.

After the data collection is complete, \textbf{(S3) preprocessing data} step involves pre-calculating any quantities used repeatedly in the training step. In our case, it is the ideal unitary corresponding to the control variables. Although not required, it is a recommended step. In our case, we pre-calculate the ideal unitary to speed up the computation.

We perform \textbf{(S4) model selection} to search for optimal model configurations within the search spaces. We use \texttt{optuna} implemented with \texttt{ray-tune} to search for optimal configurations, i.e., hyperparameter tuning. \texttt{ray-tune} can perform distributed optimization over multiple processing instances (see Appendix \ref{sec:packages} for more details about packages used in this work).
The architecture of the Blackbox consists of three group MLP layers (see \cref{fig:model-architecture} for the illustration of the model architecture). The first group is the Shared layer (MLP of 2 layers), which is shared for all Pauli observables. After processing by Shared layer, Pauli (MLP of two layers) and Hermitian (MLP of single layer) groups will independently transform the result from the Shared layer and return parameters for \cref{eq:wo} of each Pauli observables (see \cref{sec:wo-modelling} for the choice of parametrized Hermitian matrix.)
We use \texttt{relu} activation function for the model except for the output of Hermitian layers that use \texttt{tanh} and \texttt{sigmoid} activation functions to parametrized Hermitian matrix. We scaled the output to the range of $0$ to $2\pi$.
We optimize the number of neurons of each hidden layer of the multi-layer perceptron (MLP) neural network.
More specifically, we perform hyperparameter tuning for the size of the first layer of the Shared group connected to the input layer from 5 to 9 and the sizes of the 2nd - 4th layers (the rest of the remaining layers) from 5 to 49 units.
While it is possible to include other components, such as learning algorithms (optimizers) as hyperparameters, it is not within the scope of this study.

In \textbf{each model training given hyperparameter configuration}, we set our model training loop by aiming to (a) finish within a reasonable time and (b) the model is not overfitting or underfitting the training dataset.
For the training setting, we discuss our choices as follows,
\begin{itemize}
    \item We randomly split the dataset into a training, validating, and testing dataset with 800, 100, and 100 samples, respectively. Typically, we use the validation dataset for an adaptive learning rate scheduler or early stopping mechanism. We observe that it is not necessary in our case, but we keep the validation dataset within our procedure for the sake of implementation completeness. The average of $\mseeloss$ of the testing dataset is optimized by a hyperparameter tuner. We select a batch size of 100 samples, resulting in 8 iterations per epoch. We train a model for 1,500 epochs, resulting in 12,000 training iterations. Note that the numbers are arbitrary and must be adjusted for different circumstances. For example, the number of iterations should be large enough such that the metrics converge.
    \item Based on the \texttt{adam} optimizer widely used in the machine learning community, we chose \texttt{adamw} with a weight decay mechanism. We chose \texttt{warmup-cosine-decay-scheduler} predefined in \texttt{optax} \cite{deepmind2020jax} as a learning rate schedule. We refer to our GitHub for further details \cite{Pathumsoot_Inspeqtor_2025}. Our learning rate scheduler is set to actively adjust the learning rate for the first 10,000 iterations out of a total of 12,000 iterations, and then the learning rate will remain constant. The extra 2,000 iterations in which the learning rate is constant are to make sure that the model finally converges and stops learning.
    \item Our architecture of choice is a regular multi-layer perceptron (MLP). There is no obvious choice in the architecture associated with our data and context. We chose MLP as it is simple to implement and train (see Appendix \ref{sec:wo-modelling} for the parametrization of $\hat{W}_{O}$). In \cref{sec:idealshots} and \cref{sec:experiment}, we show empirically that a simple MLP is enough to implement a Graybox model that can predict the expected $\mseeloss$ approach the theoretical expected minimum value.
    \item There are two choices of loss functions considered in this work, the MSE of expectation values, $\mseeloss$ in~\cref{eq:msee}, and the AE of average gate fidelities (AGF), $\maefloss$ in~\cref{eq:maef}. These two choices arise from the discussion in \cref{sec:model-learnability} and \cref{sec:performance-bound}.
\end{itemize}

The expected minimum $\mseeloss$ obtained from our result in \cref{eq:var-data} can help researchers to select a model that accurately characterizes the device. For example, if the $\mseeloss$ predicted by the Graybox model does not approximately approach the theoretical value, one might consider adjusting the architecture of the Blackbox model, the learning rate, or other strategies that can reduce the reducible error. Suppose the expected $\mseeloss$ prediction converges higher than the theoretical value. In that case, one might measure the data variance of the dataset by an additional experiment on a real device, which becomes the experimental expected minimum value of expected $\mseeloss$.

After the model selection, we perform \textbf{(S5) model calibration} to optimize the control variables to a target quantum gate. The optimization is performed by using a gradient-based optimization algorithm. We use the \texttt{adam} optimizer with the \texttt{warmup-cosine-decay-scheduler} learning rate schedule for the active period of 1,000 iterations. This schedule is practically the same as the schedule used in training the model but is set to be active for 1,000 iterations. The bounds of control variables should be the same as those used to produce control variables in the dataset.
We perform \textbf{(S6) benchmarking} to evaluate the performance of optimized control against the target quantum gate on the real device using the AGF defined in \cref{eq:direct-agf-2}.

\subsection{Example: Superconducting qubit} \label{sec:superconducting-specific}

We note again that the procedure we presented in the previous subsection can be applied to any qubit platform. In our work, we choose to work with superconducting transmon qubits from IBM Quantum~\cite{abughanemIBMQuantumComputers2025, chowImplementingStrandScalable2014}.
It is important to note that the readout processes and qubit's state dynamics on the IBM Quantum device are defined in a rotating frame (in addition to the interaction frame in Eq.~\eqref{eq:exp-wo}) with respect to the qubit frequency $\omega_{q}$. Therefore, the ideal Hamiltonian for the transmon qubit describing the microwave drive is given by
\begin{align} \label{eq:h-rot}
    \hat{H}_{\text{ideal}}(\control,t) =\,  &( 2 \pi \Omega ) s(\mathbf{\Theta}, t) \times \nonumber \\ 
    & [\cos{ (2 \pi\omega_q t)} \hat{\sigma}_x - \sin{ (2 \pi \omega_q t)} \hat{\sigma}_y],
\end{align}
where $\Omega$ is the drive strength~\cite{javadi-abhariQuantumComputingQiskit2024}. The signal function $s(\mathbf{\Theta}, t)$ is of the form
\begin{equation}
    s(\mathbf{\Theta},t) = \text{Re} \left\{ h(\mathbf{\Theta}, t) e^{i(2 \pi \omega_d t + \phi)} \right\},
\end{equation}
describing the microwave field with an envelope function $h(\mathbf{\Theta}, t)$, a carrier frequency $\omega_{d}$ which is typically set to be on resonance with the qubit frequency, and a phase $\phi$. The phase $\phi$ is managed by \texttt{qiskit}, a python package to handle interaction with the IBMQ device, to implement a Virtual-Z gate \cite{mckayEfficient$Z$Gates2017}.

Although one can define $h(\mathbf{\Theta}, t)$ as an arbitrary function, the choice of control is directly tied to applications of interest. For example, one can perform quantum noise spectroscopy (QNS) further to extract noise information \cite{chalermpusitarakFrameBasedFilterFunctionFormalism2021} by selecting the CMPG pulse sequence. Also, the limitations in the actual implementation must be considered. In our case, (1) the function will be sampled with finite resolution, limited by each device. For \texttt{ibm\_kawasaki}, this resolution is set to $0.22$\si{ns} per sample. Thus, the function should not require resolution beyond the device's discretization capability. (2) The amplitude is limited to $1$, so the maximum amplitude should not exceed the limit.

As the target qubit is a transmon qubit, population leakage to higher energy levels is unavoidable, especially when a qubit is driven by a short pulse control \cite{krantzQuantumEngineersGuide2019}.
The way to get around this is to use the DRAG pulse technique \cite{motzoiSimplePulsesElimination2009} to help mitigate the leakage problem. This involves adding a derivative of the original pulse envelope, $\partial f(\mathbf{\Theta}, t)$, in the imaginary component, resulting in the control pulse envelope of the form,
\begin{equation}
    h(\mathbf{\Theta}, t) = f(\mathbf{\Theta}, t) + i\gamma \frac{\partial f(\mathbf{\Theta}, t)}{\partial t}.
\end{equation}
We aimed to optimize for the \sx gate as an example. Thus, the range of $\control$ was chosen such that it contains control variables of \sx gate in a noiseless condition. Since a proper sequence of \sx and Virtual-Z gates can produce a universal gateset \cite{mckayEfficient$Z$Gates2017}.

We consider the pulse envelope defined as follows. Consider a Gaussian envelope defined as $g(\sigma, c, t) = \text{exp}(\frac{-(t-c)^2}{2 \sigma^2})$, parametrized by a standard deviation $\sigma$ and a peak at $c$. A lifted Gaussian envelope with amplitude $A \in (0, 1 )$ is defined as
\begin{equation} \label{eq:lifted-gaussian-basis}
    l_{i, j} (A, \sigma, t; c) = A \frac{g(\sigma, c, t) - g(\sigma, c, -1)}{1 - g(\sigma, c, -1)},
\end{equation}
such that the first sample will be zero \cite{javadi-abhariQuantumComputingQiskit2024}. However, we want to increase the search space, so we consider the multiple-order superposition of pulse defined in \cref{eq:lifted-gaussian-basis}. We expected the first leading order to be the main driving term to the target state. At the same time, we interpret the higher-order term as a \emph{correction terms} of the total pulse shape. The concept of optimizing over the basis of pulses is also used in \cite{canevaChoppedRandombasisQuantum2011, hyyppaReducingLeakageSingleQubit2024}. Hence, we denote a set of $\zeta = \{\sigma_{i, j}\}_{i, j}$ and $\mathbf{A} = \{A_{i, j}\}_{i, j}$ for $i$ orders and $j^{\text{th}}$ pulse in the order, we defined an envelope function of the form,
\begin{equation}
    v(\mathbf{A}, \zeta, t; \mathbf{c}) = \sum_{i=1}^{4} \frac{1}{2^{i}} \left(\sum_{j=1}^{i} l_{i ,j}(A_{i, j}, \sigma_{i, j}, t;  c_{i, j}) \right).
\end{equation}
Note that we limit the number of orders to $i = 4$ so that the total number of parameters to be optimized is reasonable. The standard deviation of the basis Gaussian envelopes has different constraints depending on the order, $i$. For the $i^{\text{th}}$ order, $\sigma_i \in [a_i, b_i]$ where $(a_i, b_i)$ are chosen to be $(7,9)$, $(5,7)$, $(3,5)$, and $(1,3)$ for $i = 1,2,3,4$ respectively. The peak position $c$ of the basis envelopes depends on the order, $i$, and the term, $j$. For each $i^{\text{th}}$ order, each $j^{\text{th}}$ Gaussian pulse evenly place itself within the total pulse length of $17.78 \text{ns}$.


Finally, with DRAG technique, we have a final control with DRAG coefficient $\gamma \in (-2, 2)$, of the following form,
\begin{equation} \label{eq:control}
    h(\mathbf{A}, \zeta, \gamma, t; \mathbf{c}, t) = v(\mathbf{A}, \zeta, t; \mathbf{c}, t) + i\gamma \frac{\partial v(\mathbf{A}, \zeta, t; \mathbf{c}, t)}{\partial t},
\end{equation}
where we denote a set of control parameter $\{ \mathbf{A}, \zeta, \gamma \}$ to be sampled.

\subsection{Numerical: Data variance as the optimal point} \label{sec:idealshots}
\begin{figure}[t]
    \centering
    \includegraphics[width=1.0\linewidth]{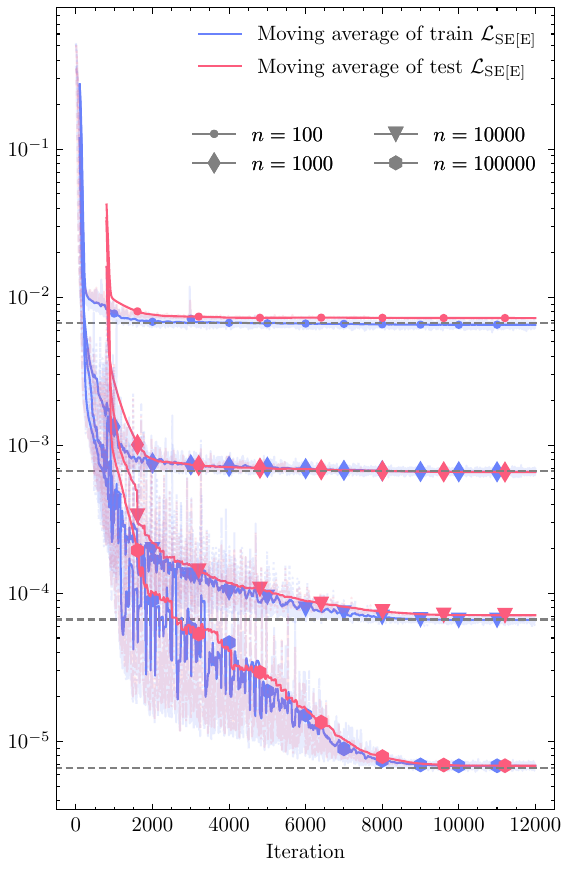}
    \caption{The train and test losses of models trained on ideal simulated data with varying numbers of shots, $n$, used to estimate expectation values. Gray dashed lines are the data variance calculated using \cref{eq:var-data} serve as the lower bound of expected $\mseeloss$ loss. The colored lines are the moving average of the gray-out line.}
    \label{fig:ideal-shots}
\end{figure}
We begin with a numerical simulation of the characterization procedure in a noiseless condition with varying finite measurement shots $n$.
Consequently, we show that the expected loss over training and testing dataset $\mathbb{E}_{\mathrm{train}}[\mseeloss] \gtrsim \frac{2}{3n}$ and $\mathbb{E}_{\mathrm{test}}[\mseeloss] \gtrsim \frac{2}{3n}$ are achievable by the models as expected. The simulation process is as follows. The ideal evolution of the system without noise is given by $\hat{H}_{\text{total}} = \hat{H}_{\text{ideal}}$, where the Hamiltonian and the control are defined in \cref{eq:h-rot} and \cref{eq:control}.
In this ideal case, it is straightforward to use the Whitebox ODE solver to solve for the ideal unitary directly given the control variables.

With the ideal unitary, the initial states $\rho_0 \in \{ |+\rangle \langle +|, |-\rangle \langle -|, |i\rangle \langle i|, |-i\rangle \langle -i|, |0\rangle \langle 0|, |1\rangle \langle 1| \}$, and the observables $\hat{P} \in \{\hat{X}, \hat{Y}, \hat{Z}\}$, we calculate the probability of observing measurement readout 1 and -1 and then sample the readouts for a given number of shots. Finally, the expectation value is calculated by averaging over all measurement readouts. Thus, the only source of uncertainty in our dataset is the finite-shot estimation of the expectation value. We numerically simulate the dataset using different numbers of shots $n \in \{10^2, 10^3, 10^4, 10^5 \}$ and train the model for each dataset with our model training routine \cref{sec:pipeline}. However, since the regression is simple enough, we randomly sample a single hyperparameter configuration for training.

The value of $\mseeloss$ throughout the training iteration is shown in \cref{fig:ideal-shots} for different numbers of measurement shots.
Since the convergence of each model is visually separated, we plot $\mseeloss^{\text{train}}$ and $\mseeloss^{\text{test}}$ of each $n$ together.
The data point associated with $\mseeloss^{\text{train}}$ is the average over a batch in the training set. On the other hand, we evaluated $\mseeloss^{\text{test}}$ once at the end of each epoch, so each data point associated with $\mseeloss^{\text{test}}$ is the average of $\mseeloss^{\text{test}}$ over the entire testing dataset. For the sake of visualization, we plot the moving average with a window size of 100 as the colored lines.
The horizontal dashed lines correspond to the data variance calculated from \cref{eq:var-data} using the $n$-shots dataset to train each model.
As shown, all random models achieve near-optimal performance of $\frac{2}{3n}$.

\subsection{Experimental: Characterization \& Calibration} \label{sec:experiment}

Equipped with the analytical tools and numerical results, we now focus on the experimental realization of the characterization and calibration procedure and benchmarking of the control on a real device.
(S1) We sample parameters of control defined in \cref{eq:control} for \samplesize samples. For each control sample, we need 18 expectation value estimations of 3 observables and 6 different initial states. This results in a total of 18,000 quantum circuits to be executed.
For each of the 18,000 quantum circuits, we use $n = 3,000$ as the number of shots.
We chose \samplesize samples so that the whole experiment is within the limitation of 1 session (8 hours) allowed by our access to the device (our experiment ran for approximately 4 hours.)
(S2) The experiment was performed on qubit \texttt{q2} of \texttt{ibm\_kawasaki}, a 127-qubit system.
(S3) The qubit specifications used in Whitebox construction are as follows, $\omega_q = 5.081\si{GHz}$ and $\Omega = 0.103\si{GHz}$.
As discussed in \cref{sec:pipeline}, we consider two loss functions, i.e., $\mseeloss$ for the general predictive model and $\maefloss$ for the AGF-specific predictive model.
We denote the models using the corresponding loss functions by $\mseemodel$ and $\maefmodel$, respectively.
(S4) Using the experimental data, we performed the model selection, resulting in the model size of 4,036 trainable parameters for $\mseemodel$ and 11,115 trainable parameters for $\maefmodel$.

\begin{figure}
    \centering
    \includegraphics[width=1\linewidth]{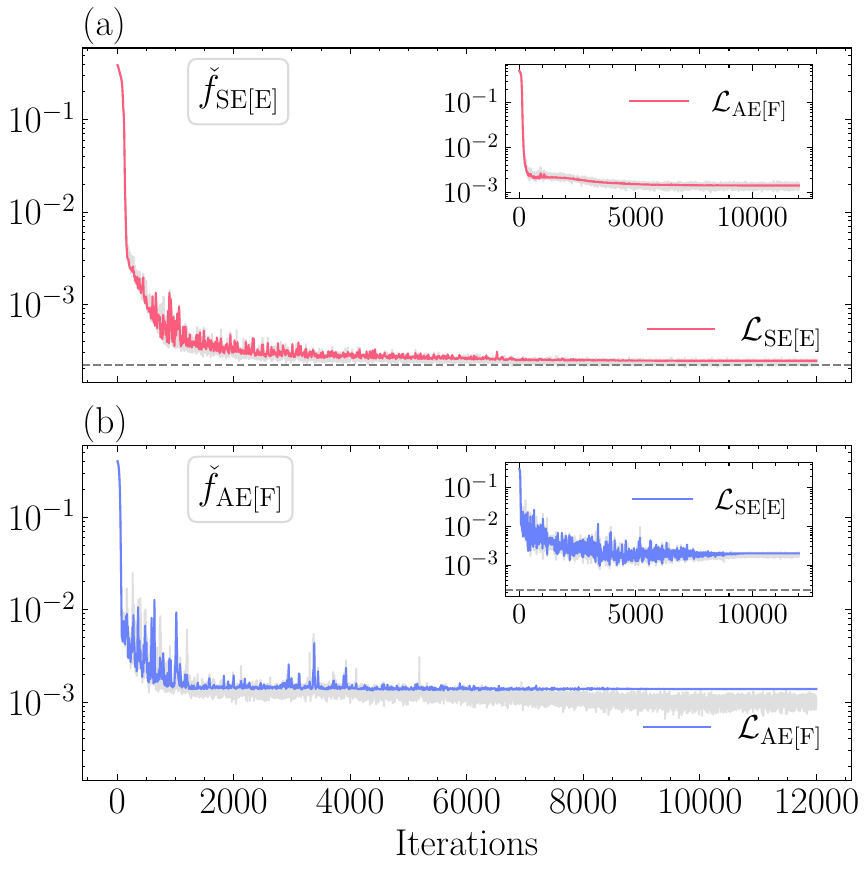}
    \caption{
        The plots of metrics over 12,000 training iterations, where the upper and lower are plots of $\mseemodel$ and $\maefmodel$, respectively.
        In (a) and (b), the vertical axis of the main plot is $\mseeloss$ ($\maefloss$), while the vertical axis of the inset plot is $\maefloss$ ($\mseeloss$).
        The colored line is the average value over the testing dataset, while the gray line is the average over a batch of the training dataset.
    }
    \label{fig:model-training}
\end{figure}

\cref{fig:model-training} shows the metrics of both models during the training phase.
We plot the value of the loss function using a testing dataset with colored lines and the value of the loss function using a training dataset with gray lines.
In the plots with $\mseeloss$ as the loss function, the gray dashed lines represent the data variance of $\frac{2}{3n}$, which is the minimum value of expected $\mseeloss$.
We observe in \cref{fig:model-training}(a) that the value of $\mseeloss$ approaches the expected minimum value. In contrast, this is not the case for $\maefmodel$, as shown in the inset of \cref{fig:model-training}(b), as expected because minimizing $\maefloss$ does not imply minimizing $\mseeloss$.
Thus, the model trained using $\maefloss$ is only good for the AGF prediction.
That is, the expectation value predicted by $\maefmodel$ is not reliable. Consequently, the function of expectation values using prediction from $\maefmodel$ rather than AGF will be inaccurate.
Furthermore, $\maefmodel$ is overfitting the training dataset as the gray line of $\maefloss$ converged to a value lower than the colored line, representing the average value of the testing dataset.

\begin{figure*}
    \subfloat[Pulse sequence using $\control_{\mathrm{SE[E]}}$]{\label{fig:MSEE_control}
        \includegraphics[width=0.49\textwidth]{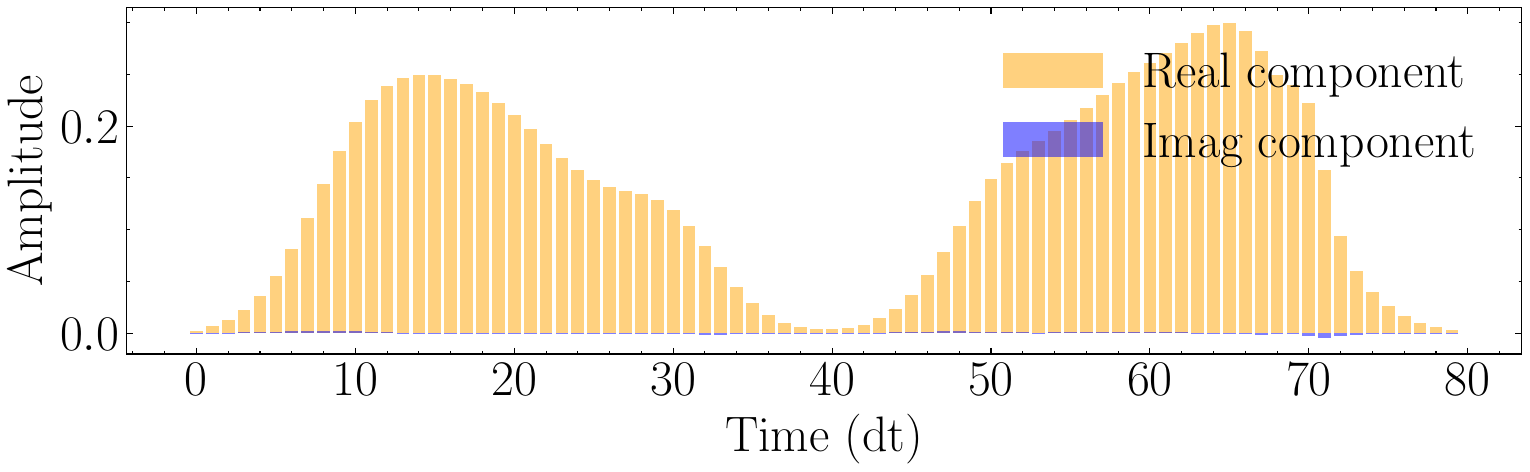}}
    \subfloat[Pulse sequence using $\control_{\mathrm{AE[F]}}$]{\label{fig:AEF_control}
        \includegraphics[width=0.49\textwidth]{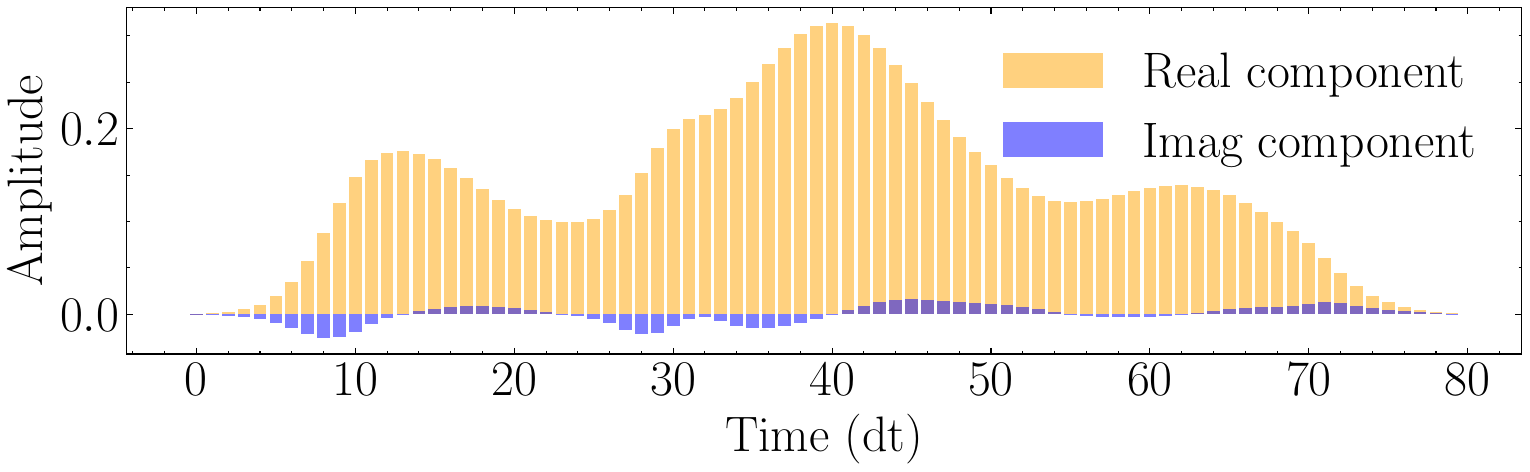}}

    \subfloat[Experimental data using $\control_{\mathrm{SE[E]}}$]{\label{fig:MSEE_benchmarking}
        \includegraphics[width=0.49\textwidth]{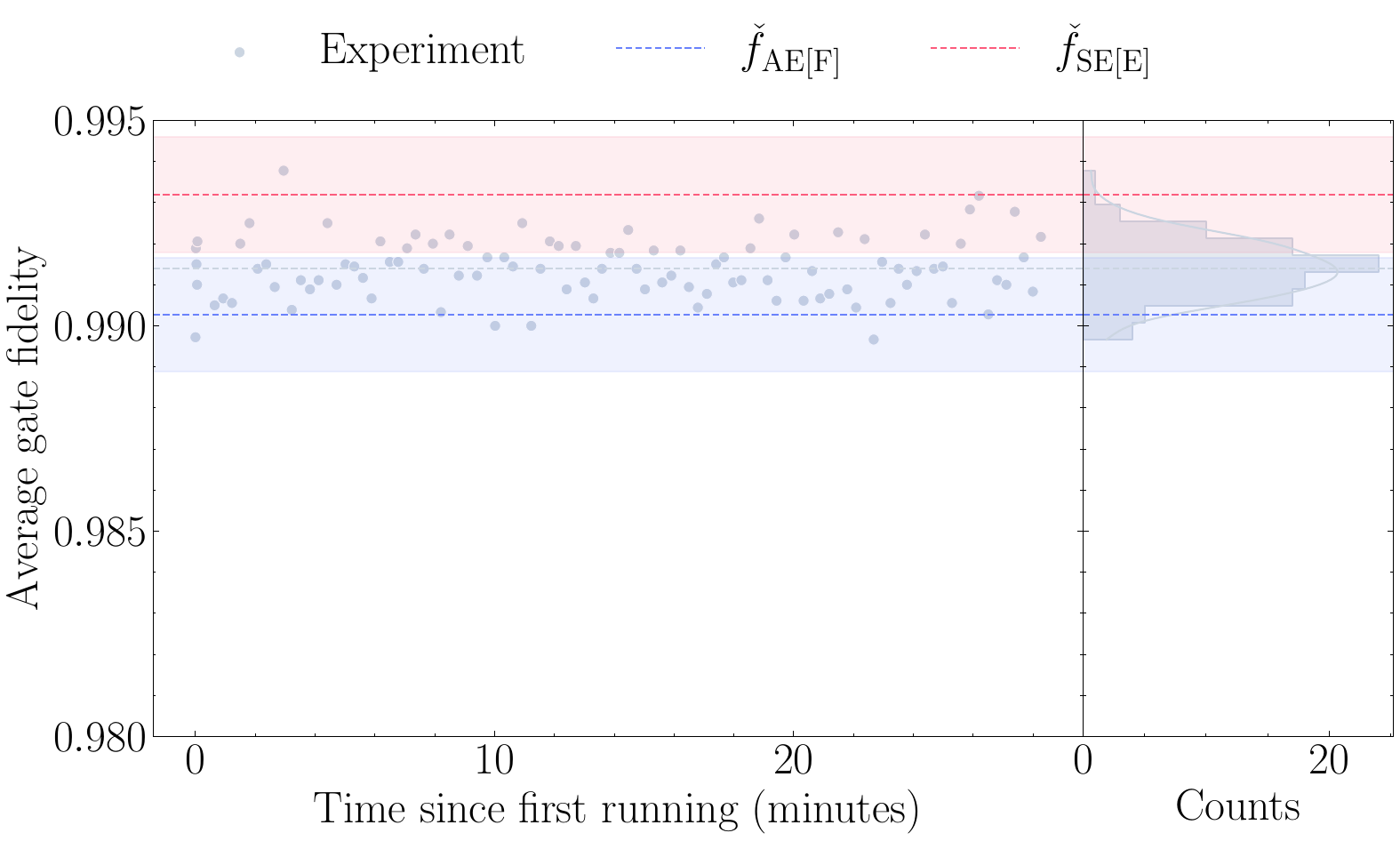}}
    \subfloat[Experimental data using $\control_{\mathrm{AE[F]}}$]{\label{fig:AEF_benchmarking}
        \includegraphics[width=0.49\textwidth]{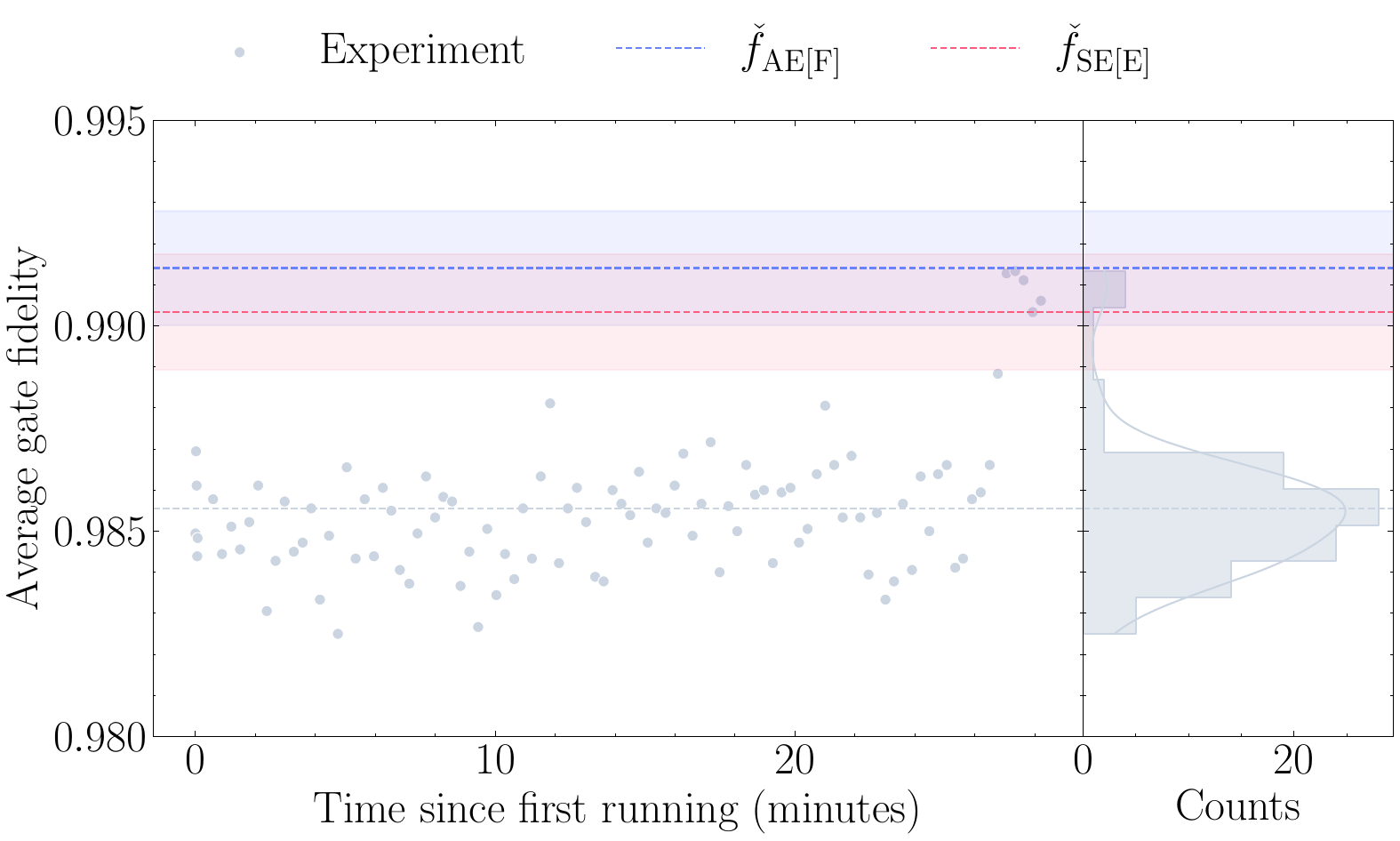}}
    \caption{
        Average gate fidelities from the benchmarking experiment performed on \texttt{ibm\_kawasaki} by the pulse sequence optimized using each model, each in the inset plot. The scatter plot on the left shows the 100 AGF against the time experiment executed on the device. On the right is the histogram of AGF data points on the left. The dashed blue line is an AGF predicted by $\maefmodel$ with uncertainty as the expected test $\maefloss$. The dashed red line is an AGF predicted by $\mseemodel$ with uncertainty as the expected test $\maefloss$.
    }
    \label{fig:model-benchmarking}
\end{figure*}

To evaluate our model's predictability and demonstrate the (S5) control calibration using the Graybox characterization approach, we optimized for the target gate, $\hat{G} = $ \sx. We use the average gate infidelity ${\cal C} = 1 - \bar{F} ( \mathcal{E},\hat{G})$ as the cost function.
We denote the controls that optimize the cost ${\cal C}$ using the trained model $\mseemodel$, and $\maefmodel$, $\control_{\mathrm{SE[E]}}$ and $\control_{\mathrm{AE[F]}}$, respectively.
The corresponding optimized pulse envelope of each model is plotted in \cref{fig:MSEE_control} and \cref{fig:AEF_control}.
(S6) Using the control pulses, we then performed direct AGF estimation 100 times consecutively on \texttt{ibm\_kawasaki} for each control parameter from the optimization using both models.
We note that the time gap between the last characterization experimental data point (finished) and the first (running) data point for $\control_{\mathrm{AE[F]}}$ benchmarking experiment is approximately 2 hours 38 minutes. Also, the time gap between the last characterization experimental data point (finished) and the first (running) data point for $\control_{\mathrm{SE[E]}}$ benchmarking experiment is approximately 3 hours 12 minutes.
The plot of AGF of each model is shown in \cref{fig:MSEE_benchmarking} and \cref{fig:AEF_benchmarking}.
Summary of the results with corresponding uncertainties is shown in~\cref{tab:avg-table}. The prediction uncertainties are calculated as follows.
(1) Average of $\maefloss$ over the testing dataset, denoted by $\mathbb{E}_{\mathrm{test}}[\maefloss]$, is the uncertainty based on the expected performance of the model when predicting AGF. We used an expected value over the testing dataset to represent the performance of the model for unseen data.
(2) The standard deviation of experimental data.

\begin{table}
    \centering
    \caption{
    The table shows the average gate fidelity (AGF) of experimental data and predicted values by $\mseemodel$ and $\maefmodel$ for control variables. The uncertainty of experimental data is a standard deviation of 100 samples. The uncertainty of prediction is $\mathbb{E}_{\mathrm{test}}[\maefloss]$.
    }
    \label{tab:avg-table}
    \begin{tabular}{lccc}
        \hline
                               & \multicolumn{2}{c}{AGF}     &                             \\
        \hline
                               & $\control_{\mathrm{SE[E]}}$ & $\control_{\mathrm{AE[F]}}$ \\
        \hline
        $\mseemodel$           & $99.3 \pm 0.1$              & $99.0 \pm 0.1$              \\
        $\maefmodel$           & $99.0 \pm 0.1$              & $99.1 \pm 0.1$              \\
        \texttt{ibm\_kawasaki} & $99.1 \pm 0.1$              & $98.6 \pm 0.2$              \\
        \hline
    \end{tabular}
\end{table}

Even with the prediction uncertainty, the experimental AGF data is not contained within the error bound, which might be due to the drift of the parameters or noise over time, as the time period for dataset collection was before the period for estimating AGF. In the case of $\maefmodel$, we observed that the model overfits the training dataset. This might be another factor that contributes to the poor predictive ability of the unseen data.

\section{Conclusion} \label{sec:conclusion}

Accurately characterizing a quantum device is a crucial step in achieving optimal control.
The Graybox characterization method allows for the decoupling of ideal and noisy evolution, enabling the building of a predictive model without assumptions about the noise by implicitly learning the noise via a neural network.
In practice, only finite-shot estimation of the dataset is achievable; therefore, the reliability of the predictive model constructed from finite-shot data is a significant concern.
We first began with the general form of the predictive model, which is the primary focus of this study. Using the model, we considered the common choice of loss function in model training, i.e., the MSE of expectation values. Using the decomposition of expected MSE loss, we derived the minimum value of the expected loss achievable by the predictive model.
The analytical results were confirmed with synthetic datasets in ideal noiseless conditions, apart from
the uncertainty from estimation of expectation values using finite measurement results. Moreover, we characterized superconducting qubit devices using the Graybox model with an MLP-based deep neural network.
Despite its simplicity, we showed that the model trained using the MSE of expectation values has achieved near-optimal predictive performance.
Following the first analytical result, we subsequently derived an upper bound for the AE of AGF between exact and predictive values in terms of the MSE of expectation values produced by the predictive model, showing that in the model trained using finite-shot dataset is sufficient to reduce different of AGF between exact and predicted value to zero. We then benchmarked trained models based on the MSE of expectation values and the AE of AGF by performing control calibration using these two models on superconducting qubit devices. Though the predicted results failed to explain the experimental data, we suspected it was caused by unknown factors such as environmental drift.

In the context of characterization for control calibration, extending the method to the two-qubit case is sufficient, as up to two-qubit control is needed for a universal gate set. Thus, generalizing the analysis to two-qubit case would be interesting future work.
For model learnability as analyzed in \cref{sec:model-learnability}, it is straightforward to calculate the average of the variance of each Pauli observable of a two-qubit system measured using \cref{eq:var-observable}.
A similar procedure could be applied to derive a bound for the AGF prediction for a two-qubit gate. First, the $\mseeloss$ should be defined for two qubits: averages over expectation values. Second, instead of the single-qubit case AGF, a two-qubit AGF would need to be derived. Finally, one could substitute them into the Cauchy-Schwarz inequality for the bound of the two-qubit case.
Furthermore, extending the analysis to address the state preparation and measurement error problem is also an interesting and valuable future work.

Our results provided researchers with additional information to make an even more informed decision in experimentation, saving the cost of trial and error on a real device.

\section*{Acknowledgment}

We would like to thank Thiparat Chotibut, Sujin Suwanna, Behnam Tonekaboni, Kaiah Steven, and Gerardo Paz-Silva for fruitful discussions. We acknowledge the use of IBM Quantum services for this work. The views expressed are those of the authors, and do not reflect the official policy or position of IBM or the IBM Quantum team. This work was supported by JST [Moonshot R\&D Program] Grant
Number [JPMJMS226C] and [JPMJMS2061]. A.~C.~acknowledges the support of the Program Management Unit for Human Resources and Institutional Development, Research and Innovation (Thailand) grant B39G680007.

\bibliography{references}

\begin{thebibliography}{37}%
\makeatletter
\providecommand \@ifxundefined [1]{%
 \@ifx{#1\undefined}
}%
\providecommand \@ifnum [1]{%
 \ifnum #1\expandafter \@firstoftwo
 \else \expandafter \@secondoftwo
 \fi
}%
\providecommand \@ifx [1]{%
 \ifx #1\expandafter \@firstoftwo
 \else \expandafter \@secondoftwo
 \fi
}%
\providecommand \natexlab [1]{#1}%
\providecommand \enquote  [1]{``#1''}%
\providecommand \bibnamefont  [1]{#1}%
\providecommand \bibfnamefont [1]{#1}%
\providecommand \citenamefont [1]{#1}%
\providecommand \href@noop [0]{\@secondoftwo}%
\providecommand \href [0]{\begingroup \@sanitize@url \@href}%
\providecommand \@href[1]{\@@startlink{#1}\@@href}%
\providecommand \@@href[1]{\endgroup#1\@@endlink}%
\providecommand \@sanitize@url [0]{\catcode `\\12\catcode `\$12\catcode `\&12\catcode `\#12\catcode `\^12\catcode `\_12\catcode `\%12\relax}%
\providecommand \@@startlink[1]{}%
\providecommand \@@endlink[0]{}%
\providecommand \url  [0]{\begingroup\@sanitize@url \@url }%
\providecommand \@url [1]{\endgroup\@href {#1}{\urlprefix }}%
\providecommand \urlprefix  [0]{URL }%
\providecommand \Eprint [0]{\href }%
\providecommand \doibase [0]{https://doi.org/}%
\providecommand \selectlanguage [0]{\@gobble}%
\providecommand \bibinfo  [0]{\@secondoftwo}%
\providecommand \bibfield  [0]{\@secondoftwo}%
\providecommand \translation [1]{[#1]}%
\providecommand \BibitemOpen [0]{}%
\providecommand \bibitemStop [0]{}%
\providecommand \bibitemNoStop [0]{.\EOS\space}%
\providecommand \EOS [0]{\spacefactor3000\relax}%
\providecommand \BibitemShut  [1]{\csname bibitem#1\endcsname}%
\let\auto@bib@innerbib\@empty
\bibitem [{\citenamefont {Suchara}\ \emph {et~al.}(2013)\citenamefont {Suchara}, \citenamefont {Kubiatowicz}, \citenamefont {Faruque}, \citenamefont {Chong}, \citenamefont {Lai},\ and\ \citenamefont {Paz}}]{sucharaQuREQuantumResource2013}%
  \BibitemOpen
  \bibfield  {author} {\bibinfo {author} {\bibfnamefont {M.}~\bibnamefont {Suchara}}, \bibinfo {author} {\bibfnamefont {J.}~\bibnamefont {Kubiatowicz}}, \bibinfo {author} {\bibfnamefont {A.}~\bibnamefont {Faruque}}, \bibinfo {author} {\bibfnamefont {F.~T.}\ \bibnamefont {Chong}}, \bibinfo {author} {\bibfnamefont {C.-Y.}\ \bibnamefont {Lai}},\ and\ \bibinfo {author} {\bibfnamefont {G.}~\bibnamefont {Paz}},\ }\bibfield  {title} {\bibinfo {title} {{{QuRE}}: {{The Quantum Resource Estimator}} toolbox},\ }in\ \href {https://doi.org/10.1109/ICCD.2013.6657074} {\emph {\bibinfo {booktitle} {2013 {{IEEE}} 31st {{International Conference}} on {{Computer Design}} ({{ICCD}})}}}\ (\bibinfo {year} {2013})\ pp.\ \bibinfo {pages} {419--426}\BibitemShut {NoStop}%
\bibitem [{\citenamefont {Bravyi}\ \emph {et~al.}(2024)\citenamefont {Bravyi}, \citenamefont {Cross}, \citenamefont {Gambetta}, \citenamefont {Maslov}, \citenamefont {Rall},\ and\ \citenamefont {Yoder}}]{bravyiHighthresholdLowoverheadFaulttolerant2024}%
  \BibitemOpen
  \bibfield  {author} {\bibinfo {author} {\bibfnamefont {S.}~\bibnamefont {Bravyi}}, \bibinfo {author} {\bibfnamefont {A.~W.}\ \bibnamefont {Cross}}, \bibinfo {author} {\bibfnamefont {J.~M.}\ \bibnamefont {Gambetta}}, \bibinfo {author} {\bibfnamefont {D.}~\bibnamefont {Maslov}}, \bibinfo {author} {\bibfnamefont {P.}~\bibnamefont {Rall}},\ and\ \bibinfo {author} {\bibfnamefont {T.~J.}\ \bibnamefont {Yoder}},\ }\bibfield  {title} {\bibinfo {title} {High-threshold and low-overhead fault-tolerant quantum memory},\ }\href {https://doi.org/10.1038/s41586-024-07107-7} {\bibfield  {journal} {\bibinfo  {journal} {Nature}\ }\textbf {\bibinfo {volume} {627}},\ \bibinfo {pages} {778} (\bibinfo {year} {2024})}\BibitemShut {NoStop}%
\bibitem [{\citenamefont {Baum}\ \emph {et~al.}(2021)\citenamefont {Baum}, \citenamefont {Amico}, \citenamefont {Howell}, \citenamefont {Hush}, \citenamefont {Liuzzi}, \citenamefont {Mundada}, \citenamefont {Merkh}, \citenamefont {Carvalho},\ and\ \citenamefont {Biercuk}}]{baumExperimentalDeepReinforcement2021}%
  \BibitemOpen
  \bibfield  {author} {\bibinfo {author} {\bibfnamefont {Y.}~\bibnamefont {Baum}}, \bibinfo {author} {\bibfnamefont {M.}~\bibnamefont {Amico}}, \bibinfo {author} {\bibfnamefont {S.}~\bibnamefont {Howell}}, \bibinfo {author} {\bibfnamefont {M.}~\bibnamefont {Hush}}, \bibinfo {author} {\bibfnamefont {M.}~\bibnamefont {Liuzzi}}, \bibinfo {author} {\bibfnamefont {P.}~\bibnamefont {Mundada}}, \bibinfo {author} {\bibfnamefont {T.}~\bibnamefont {Merkh}}, \bibinfo {author} {\bibfnamefont {A.~R.}\ \bibnamefont {Carvalho}},\ and\ \bibinfo {author} {\bibfnamefont {M.~J.}\ \bibnamefont {Biercuk}},\ }\bibfield  {title} {\bibinfo {title} {Experimental {{Deep Reinforcement Learning}} for {{Error-Robust Gate-Set Design}} on a {{Superconducting Quantum Computer}}},\ }\href {https://doi.org/10.1103/PRXQuantum.2.040324} {\bibfield  {journal} {\bibinfo  {journal} {PRX Quantum}\ }\textbf {\bibinfo {volume} {2}},\ \bibinfo {pages} {040324} (\bibinfo {year} {2021})}\BibitemShut {NoStop}%
\bibitem [{\citenamefont {Werninghaus}\ \emph {et~al.}(2021)\citenamefont {Werninghaus}, \citenamefont {Egger}, \citenamefont {Roy}, \citenamefont {Machnes}, \citenamefont {Wilhelm},\ and\ \citenamefont {Filipp}}]{werninghausLeakageReductionFast2021}%
  \BibitemOpen
  \bibfield  {author} {\bibinfo {author} {\bibfnamefont {M.}~\bibnamefont {Werninghaus}}, \bibinfo {author} {\bibfnamefont {D.~J.}\ \bibnamefont {Egger}}, \bibinfo {author} {\bibfnamefont {F.}~\bibnamefont {Roy}}, \bibinfo {author} {\bibfnamefont {S.}~\bibnamefont {Machnes}}, \bibinfo {author} {\bibfnamefont {F.~K.}\ \bibnamefont {Wilhelm}},\ and\ \bibinfo {author} {\bibfnamefont {S.}~\bibnamefont {Filipp}},\ }\bibfield  {title} {\bibinfo {title} {Leakage reduction in fast superconducting qubit gates via optimal control},\ }\href {https://doi.org/10.1038/s41534-020-00346-2} {\bibfield  {journal} {\bibinfo  {journal} {npj Quantum Information}\ }\textbf {\bibinfo {volume} {7}},\ \bibinfo {pages} {14} (\bibinfo {year} {2021})}\BibitemShut {NoStop}%
\bibitem [{\citenamefont {Caneva}\ \emph {et~al.}(2011)\citenamefont {Caneva}, \citenamefont {Calarco},\ and\ \citenamefont {Montangero}}]{canevaChoppedRandombasisQuantum2011}%
  \BibitemOpen
  \bibfield  {author} {\bibinfo {author} {\bibfnamefont {T.}~\bibnamefont {Caneva}}, \bibinfo {author} {\bibfnamefont {T.}~\bibnamefont {Calarco}},\ and\ \bibinfo {author} {\bibfnamefont {S.}~\bibnamefont {Montangero}},\ }\bibfield  {title} {\bibinfo {title} {Chopped random-basis quantum optimization},\ }\href {https://doi.org/10.1103/PhysRevA.84.022326} {\bibfield  {journal} {\bibinfo  {journal} {Physical Review A}\ }\textbf {\bibinfo {volume} {84}},\ \bibinfo {pages} {022326} (\bibinfo {year} {2011})}\BibitemShut {NoStop}%
\bibitem [{\citenamefont {Khalid}\ \emph {et~al.}(2023)\citenamefont {Khalid}, \citenamefont {Weidner}, \citenamefont {Jonckheere}, \citenamefont {Schirmer},\ and\ \citenamefont {Langbein}}]{khalidSampleefficientModelbasedReinforcement2023}%
  \BibitemOpen
  \bibfield  {author} {\bibinfo {author} {\bibfnamefont {I.}~\bibnamefont {Khalid}}, \bibinfo {author} {\bibfnamefont {C.~A.}\ \bibnamefont {Weidner}}, \bibinfo {author} {\bibfnamefont {E.~A.}\ \bibnamefont {Jonckheere}}, \bibinfo {author} {\bibfnamefont {S.~G.}\ \bibnamefont {Schirmer}},\ and\ \bibinfo {author} {\bibfnamefont {F.~C.}\ \bibnamefont {Langbein}},\ }\bibfield  {title} {\bibinfo {title} {Sample-efficient model-based reinforcement learning for quantum control},\ }\href {https://doi.org/10.1103/PhysRevResearch.5.043002} {\bibfield  {journal} {\bibinfo  {journal} {Physical Review Research}\ }\textbf {\bibinfo {volume} {5}},\ \bibinfo {pages} {043002} (\bibinfo {year} {2023})}\BibitemShut {NoStop}%
\bibitem [{\citenamefont {Koch}\ \emph {et~al.}(2022)\citenamefont {Koch}, \citenamefont {Boscain}, \citenamefont {Calarco}, \citenamefont {Dirr}, \citenamefont {Filipp}, \citenamefont {Glaser}, \citenamefont {Kosloff}, \citenamefont {Montangero}, \citenamefont {{Schulte-Herbr{\"u}ggen}}, \citenamefont {Sugny},\ and\ \citenamefont {Wilhelm}}]{kochQuantumOptimalControl2022}%
  \BibitemOpen
  \bibfield  {author} {\bibinfo {author} {\bibfnamefont {C.~P.}\ \bibnamefont {Koch}}, \bibinfo {author} {\bibfnamefont {U.}~\bibnamefont {Boscain}}, \bibinfo {author} {\bibfnamefont {T.}~\bibnamefont {Calarco}}, \bibinfo {author} {\bibfnamefont {G.}~\bibnamefont {Dirr}}, \bibinfo {author} {\bibfnamefont {S.}~\bibnamefont {Filipp}}, \bibinfo {author} {\bibfnamefont {S.~J.}\ \bibnamefont {Glaser}}, \bibinfo {author} {\bibfnamefont {R.}~\bibnamefont {Kosloff}}, \bibinfo {author} {\bibfnamefont {S.}~\bibnamefont {Montangero}}, \bibinfo {author} {\bibfnamefont {T.}~\bibnamefont {{Schulte-Herbr{\"u}ggen}}}, \bibinfo {author} {\bibfnamefont {D.}~\bibnamefont {Sugny}},\ and\ \bibinfo {author} {\bibfnamefont {F.~K.}\ \bibnamefont {Wilhelm}},\ }\bibfield  {title} {\bibinfo {title} {Quantum optimal control in quantum technologies. {{Strategic}} report on current status, visions and goals for research in {{Europe}}},\ }\href {https://doi.org/10.1140/epjqt/s40507-022-00138-x} {\bibfield  {journal} {\bibinfo  {journal} {EPJ Quantum Technology}\ }\textbf {\bibinfo {volume} {9}},\ \bibinfo {pages} {19} (\bibinfo {year} {2022})}\BibitemShut {NoStop}%
\bibitem [{\citenamefont {Dong}\ and\ \citenamefont {Petersen}(2010)}]{dongQuantumControlTheory2010}%
  \BibitemOpen
  \bibfield  {author} {\bibinfo {author} {\bibfnamefont {D.}~\bibnamefont {Dong}}\ and\ \bibinfo {author} {\bibfnamefont {I.}~\bibnamefont {Petersen}},\ }\bibfield  {title} {\bibinfo {title} {Quantum control theory and applications: A survey},\ }\href {https://doi.org/10.1049/iet-cta.2009.0508} {\bibfield  {journal} {\bibinfo  {journal} {IET Control Theory \& Applications}\ }\textbf {\bibinfo {volume} {4}},\ \bibinfo {pages} {2651} (\bibinfo {year} {2010})}\BibitemShut {NoStop}%
\bibitem [{\citenamefont {Ansel}\ \emph {et~al.}(2024)\citenamefont {Ansel}, \citenamefont {Dionis}, \citenamefont {Arrouas}, \citenamefont {Peaudecerf}, \citenamefont {Gu{\'e}rin}, \citenamefont {{Gu{\'e}ry-Odelin}},\ and\ \citenamefont {Sugny}}]{anselIntroductionTheoreticalExperimental2024}%
  \BibitemOpen
  \bibfield  {author} {\bibinfo {author} {\bibfnamefont {Q.}~\bibnamefont {Ansel}}, \bibinfo {author} {\bibfnamefont {E.}~\bibnamefont {Dionis}}, \bibinfo {author} {\bibfnamefont {F.}~\bibnamefont {Arrouas}}, \bibinfo {author} {\bibfnamefont {B.}~\bibnamefont {Peaudecerf}}, \bibinfo {author} {\bibfnamefont {S.}~\bibnamefont {Gu{\'e}rin}}, \bibinfo {author} {\bibfnamefont {D.}~\bibnamefont {{Gu{\'e}ry-Odelin}}},\ and\ \bibinfo {author} {\bibfnamefont {D.}~\bibnamefont {Sugny}},\ }\bibfield  {title} {\bibinfo {title} {Introduction to theoretical and experimental aspects of quantum optimal control},\ }\href {https://doi.org/10.1088/1361-6455/ad46a5} {\bibfield  {journal} {\bibinfo  {journal} {Journal of Physics B: Atomic, Molecular and Optical Physics}\ }\textbf {\bibinfo {volume} {57}},\ \bibinfo {pages} {133001} (\bibinfo {year} {2024})}\BibitemShut {NoStop}%
\bibitem [{\citenamefont {Omanakuttan}\ \emph {et~al.}(2023)\citenamefont {Omanakuttan}, \citenamefont {Mitra}, \citenamefont {Meier}, \citenamefont {Martin},\ and\ \citenamefont {Deutsch}}]{omanakuttanQuditEntanglersUsing2023}%
  \BibitemOpen
  \bibfield  {author} {\bibinfo {author} {\bibfnamefont {S.}~\bibnamefont {Omanakuttan}}, \bibinfo {author} {\bibfnamefont {A.}~\bibnamefont {Mitra}}, \bibinfo {author} {\bibfnamefont {E.~J.}\ \bibnamefont {Meier}}, \bibinfo {author} {\bibfnamefont {M.~J.}\ \bibnamefont {Martin}},\ and\ \bibinfo {author} {\bibfnamefont {I.~H.}\ \bibnamefont {Deutsch}},\ }\bibfield  {title} {\bibinfo {title} {Qudit {{Entanglers Using Quantum Optimal Control}}},\ }\href {https://doi.org/10.1103/PRXQuantum.4.040333} {\bibfield  {journal} {\bibinfo  {journal} {PRX Quantum}\ }\textbf {\bibinfo {volume} {4}},\ \bibinfo {pages} {040333} (\bibinfo {year} {2023})}\BibitemShut {NoStop}%
\bibitem [{\citenamefont {Khaneja}\ \emph {et~al.}(2005)\citenamefont {Khaneja}, \citenamefont {Reiss}, \citenamefont {Kehlet}, \citenamefont {{Schulte-Herbr{\"u}ggen}},\ and\ \citenamefont {Glaser}}]{khanejaOptimalControlCoupled2005}%
  \BibitemOpen
  \bibfield  {author} {\bibinfo {author} {\bibfnamefont {N.}~\bibnamefont {Khaneja}}, \bibinfo {author} {\bibfnamefont {T.}~\bibnamefont {Reiss}}, \bibinfo {author} {\bibfnamefont {C.}~\bibnamefont {Kehlet}}, \bibinfo {author} {\bibfnamefont {T.}~\bibnamefont {{Schulte-Herbr{\"u}ggen}}},\ and\ \bibinfo {author} {\bibfnamefont {S.~J.}\ \bibnamefont {Glaser}},\ }\bibfield  {title} {\bibinfo {title} {Optimal control of coupled spin dynamics: Design of {{NMR}} pulse sequences by gradient ascent algorithms},\ }\href {https://doi.org/10.1016/j.jmr.2004.11.004} {\bibfield  {journal} {\bibinfo  {journal} {Journal of Magnetic Resonance}\ }\textbf {\bibinfo {volume} {172}},\ \bibinfo {pages} {296} (\bibinfo {year} {2005})}\BibitemShut {NoStop}%
\bibitem [{\citenamefont {Machnes}\ \emph {et~al.}(2018)\citenamefont {Machnes}, \citenamefont {Ass{\'e}mat}, \citenamefont {Tannor},\ and\ \citenamefont {Wilhelm}}]{machnesTunableFlexibleEfficient2018}%
  \BibitemOpen
  \bibfield  {author} {\bibinfo {author} {\bibfnamefont {S.}~\bibnamefont {Machnes}}, \bibinfo {author} {\bibfnamefont {E.}~\bibnamefont {Ass{\'e}mat}}, \bibinfo {author} {\bibfnamefont {D.}~\bibnamefont {Tannor}},\ and\ \bibinfo {author} {\bibfnamefont {F.~K.}\ \bibnamefont {Wilhelm}},\ }\bibfield  {title} {\bibinfo {title} {Tunable, {{Flexible}}, and {{Efficient Optimization}} of {{Control Pulses}} for {{Practical Qubits}}},\ }\href {https://doi.org/10.1103/PhysRevLett.120.150401} {\bibfield  {journal} {\bibinfo  {journal} {Physical Review Letters}\ }\textbf {\bibinfo {volume} {120}},\ \bibinfo {pages} {150401} (\bibinfo {year} {2018})}\BibitemShut {NoStop}%
\bibitem [{\citenamefont {Fyrillas}\ \emph {et~al.}(2024)\citenamefont {Fyrillas}, \citenamefont {Faure}, \citenamefont {Maring}, \citenamefont {Senellart},\ and\ \citenamefont {Belabas}}]{fyrillasScalableMachineLearningassisted2024}%
  \BibitemOpen
  \bibfield  {author} {\bibinfo {author} {\bibfnamefont {A.}~\bibnamefont {Fyrillas}}, \bibinfo {author} {\bibfnamefont {O.}~\bibnamefont {Faure}}, \bibinfo {author} {\bibfnamefont {N.}~\bibnamefont {Maring}}, \bibinfo {author} {\bibfnamefont {J.}~\bibnamefont {Senellart}},\ and\ \bibinfo {author} {\bibfnamefont {N.}~\bibnamefont {Belabas}},\ }\bibfield  {title} {\bibinfo {title} {Scalable machine learning-assisted clear-box characterization for optimally controlled photonic circuits},\ }\href {https://doi.org/10.1364/OPTICA.512148} {\bibfield  {journal} {\bibinfo  {journal} {Optica}\ }\textbf {\bibinfo {volume} {11}},\ \bibinfo {pages} {427} (\bibinfo {year} {2024})}\BibitemShut {NoStop}%
\bibitem [{\citenamefont {Youssry}\ \emph {et~al.}(2020{\natexlab{a}})\citenamefont {Youssry}, \citenamefont {{Paz-Silva}},\ and\ \citenamefont {Ferrie}}]{youssryCharacterizationControlOpen2020}%
  \BibitemOpen
  \bibfield  {author} {\bibinfo {author} {\bibfnamefont {A.}~\bibnamefont {Youssry}}, \bibinfo {author} {\bibfnamefont {G.~A.}\ \bibnamefont {{Paz-Silva}}},\ and\ \bibinfo {author} {\bibfnamefont {C.}~\bibnamefont {Ferrie}},\ }\bibfield  {title} {\bibinfo {title} {Characterization and control of open quantum systems beyond quantum noise spectroscopy},\ }\href {https://doi.org/10.1038/s41534-020-00332-8} {\bibfield  {journal} {\bibinfo  {journal} {npj Quantum Information}\ }\textbf {\bibinfo {volume} {6}},\ \bibinfo {pages} {95} (\bibinfo {year} {2020}{\natexlab{a}})}\BibitemShut {NoStop}%
\bibitem [{\citenamefont {Youssry}\ \emph {et~al.}(2024)\citenamefont {Youssry}, \citenamefont {Yang}, \citenamefont {Chapman}, \citenamefont {Haylock}, \citenamefont {Lenzini}, \citenamefont {Lobino},\ and\ \citenamefont {Peruzzo}}]{youssryExperimentalGrayboxQuantum2024}%
  \BibitemOpen
  \bibfield  {author} {\bibinfo {author} {\bibfnamefont {A.}~\bibnamefont {Youssry}}, \bibinfo {author} {\bibfnamefont {Y.}~\bibnamefont {Yang}}, \bibinfo {author} {\bibfnamefont {R.~J.}\ \bibnamefont {Chapman}}, \bibinfo {author} {\bibfnamefont {B.}~\bibnamefont {Haylock}}, \bibinfo {author} {\bibfnamefont {F.}~\bibnamefont {Lenzini}}, \bibinfo {author} {\bibfnamefont {M.}~\bibnamefont {Lobino}},\ and\ \bibinfo {author} {\bibfnamefont {A.}~\bibnamefont {Peruzzo}},\ }\bibfield  {title} {\bibinfo {title} {Experimental graybox quantum system identification and control},\ }\href {https://doi.org/10.1038/s41534-023-00795-5} {\bibfield  {journal} {\bibinfo  {journal} {npj Quantum Information}\ }\textbf {\bibinfo {volume} {10}},\ \bibinfo {pages} {9} (\bibinfo {year} {2024})}\BibitemShut {NoStop}%
\bibitem [{\citenamefont {Youssry}\ \emph {et~al.}(2020{\natexlab{b}})\citenamefont {Youssry}, \citenamefont {Chapman}, \citenamefont {Peruzzo}, \citenamefont {Ferrie},\ and\ \citenamefont {Tomamichel}}]{youssryModelingControlReconfigurable2020}%
  \BibitemOpen
  \bibfield  {author} {\bibinfo {author} {\bibfnamefont {A.}~\bibnamefont {Youssry}}, \bibinfo {author} {\bibfnamefont {R.~J.}\ \bibnamefont {Chapman}}, \bibinfo {author} {\bibfnamefont {A.}~\bibnamefont {Peruzzo}}, \bibinfo {author} {\bibfnamefont {C.}~\bibnamefont {Ferrie}},\ and\ \bibinfo {author} {\bibfnamefont {M.}~\bibnamefont {Tomamichel}},\ }\bibfield  {title} {\bibinfo {title} {Modeling and control of a reconfigurable photonic circuit using deep learning},\ }\href {https://doi.org/10.1088/2058-9565/ab60de} {\bibfield  {journal} {\bibinfo  {journal} {Quantum Science and Technology}\ }\textbf {\bibinfo {volume} {5}},\ \bibinfo {pages} {025001} (\bibinfo {year} {2020}{\natexlab{b}})}\BibitemShut {NoStop}%
\bibitem [{\citenamefont {Auza}\ \emph {et~al.}(2024)\citenamefont {Auza}, \citenamefont {Youssry}, \citenamefont {{Paz-Silva}},\ and\ \citenamefont {Peruzzo}}]{auzaQuantumControlPresence2024}%
  \BibitemOpen
  \bibfield  {author} {\bibinfo {author} {\bibfnamefont {A.}~\bibnamefont {Auza}}, \bibinfo {author} {\bibfnamefont {A.}~\bibnamefont {Youssry}}, \bibinfo {author} {\bibfnamefont {G.}~\bibnamefont {{Paz-Silva}}},\ and\ \bibinfo {author} {\bibfnamefont {A.}~\bibnamefont {Peruzzo}},\ }\href {https://doi.org/10.48550/arXiv.2404.19251} {\bibinfo {title} {Quantum control in the presence of strongly coupled non-{{Markovian}} noise}} (\bibinfo {year} {2024}),\ \Eprint {https://arxiv.org/abs/2404.19251} {arXiv:2404.19251 [quant-ph]} \BibitemShut {NoStop}%
\bibitem [{\citenamefont {Mayevsky}\ \emph {et~al.}(2025)\citenamefont {Mayevsky}, \citenamefont {Youssry}, \citenamefont {Sareen}, \citenamefont {{Paz-Silva}},\ and\ \citenamefont {Peruzzo}}]{mayevskyQuantumEngineeringQudits2025}%
  \BibitemOpen
  \bibfield  {author} {\bibinfo {author} {\bibfnamefont {Y.}~\bibnamefont {Mayevsky}}, \bibinfo {author} {\bibfnamefont {A.}~\bibnamefont {Youssry}}, \bibinfo {author} {\bibfnamefont {R.}~\bibnamefont {Sareen}}, \bibinfo {author} {\bibfnamefont {G.~A.}\ \bibnamefont {{Paz-Silva}}},\ and\ \bibinfo {author} {\bibfnamefont {A.}~\bibnamefont {Peruzzo}},\ }\href {https://doi.org/10.48550/arXiv.2506.13075} {\bibinfo {title} {Quantum {{Engineering}} of {{Qudits}} with {{Interpretable Machine Learning}}}} (\bibinfo {year} {2025}),\ \Eprint {https://arxiv.org/abs/2506.13075} {arXiv:2506.13075 [quant-ph]} \BibitemShut {NoStop}%
\bibitem [{\citenamefont {Motzoi}\ \emph {et~al.}(2009)\citenamefont {Motzoi}, \citenamefont {Gambetta}, \citenamefont {Rebentrost},\ and\ \citenamefont {Wilhelm}}]{motzoiSimplePulsesElimination2009}%
  \BibitemOpen
  \bibfield  {author} {\bibinfo {author} {\bibfnamefont {F.}~\bibnamefont {Motzoi}}, \bibinfo {author} {\bibfnamefont {J.~M.}\ \bibnamefont {Gambetta}}, \bibinfo {author} {\bibfnamefont {P.}~\bibnamefont {Rebentrost}},\ and\ \bibinfo {author} {\bibfnamefont {F.~K.}\ \bibnamefont {Wilhelm}},\ }\bibfield  {title} {\bibinfo {title} {Simple {{Pulses}} for {{Elimination}} of {{Leakage}} in {{Weakly Nonlinear Qubits}}},\ }\href {https://doi.org/10.1103/PhysRevLett.103.110501} {\bibfield  {journal} {\bibinfo  {journal} {Physical Review Letters}\ }\textbf {\bibinfo {volume} {103}},\ \bibinfo {pages} {110501} (\bibinfo {year} {2009})}\BibitemShut {NoStop}%
\bibitem [{\citenamefont {Genois}\ \emph {et~al.}(2024)\citenamefont {Genois}, \citenamefont {Stevenson}, \citenamefont {Goss}, \citenamefont {Siddiqi},\ and\ \citenamefont {Blais}}]{genoisQuantumOptimalControl2024}%
  \BibitemOpen
  \bibfield  {author} {\bibinfo {author} {\bibfnamefont {E.}~\bibnamefont {Genois}}, \bibinfo {author} {\bibfnamefont {N.~J.}\ \bibnamefont {Stevenson}}, \bibinfo {author} {\bibfnamefont {N.}~\bibnamefont {Goss}}, \bibinfo {author} {\bibfnamefont {I.}~\bibnamefont {Siddiqi}},\ and\ \bibinfo {author} {\bibfnamefont {A.}~\bibnamefont {Blais}},\ }\href {https://doi.org/10.48550/arXiv.2410.22603} {\bibinfo {title} {Quantum optimal control of superconducting qubits based on machine-learning characterization}} (\bibinfo {year} {2024}),\ \Eprint {https://arxiv.org/abs/2410.22603} {arXiv:2410.22603 [quant-ph]} \BibitemShut {NoStop}%
\bibitem [{\citenamefont {Youssry}\ and\ \citenamefont {Nurdin}(2023)}]{youssryMultiaxisControlQubit2023}%
  \BibitemOpen
  \bibfield  {author} {\bibinfo {author} {\bibfnamefont {A.}~\bibnamefont {Youssry}}\ and\ \bibinfo {author} {\bibfnamefont {H.~I.}\ \bibnamefont {Nurdin}},\ }\bibfield  {title} {\bibinfo {title} {Multi-axis control of a qubit in the presence of unknown non-{{Markovian}} quantum noise},\ }\href {https://doi.org/10.1088/2058-9565/aca711} {\bibfield  {journal} {\bibinfo  {journal} {Quantum Science and Technology}\ }\textbf {\bibinfo {volume} {8}},\ \bibinfo {pages} {015018} (\bibinfo {year} {2023})}\BibitemShut {NoStop}%
\bibitem [{\citenamefont {Bishop}(2006)}]{bishopPatternRecognitionMachine2006}%
  \BibitemOpen
  \bibfield  {author} {\bibinfo {author} {\bibfnamefont {C.~M.}\ \bibnamefont {Bishop}},\ }\href@noop {} {\emph {\bibinfo {title} {Pattern Recognition and Machine Learning}}},\ Information Science and Statistics\ (\bibinfo  {publisher} {Springer},\ \bibinfo {address} {New York},\ \bibinfo {year} {2006})\BibitemShut {NoStop}%
\bibitem [{\citenamefont {Nielsen}(2002)}]{nielsenSimpleFormulaAverage2002}%
  \BibitemOpen
  \bibfield  {author} {\bibinfo {author} {\bibfnamefont {M.~A.}\ \bibnamefont {Nielsen}},\ }\bibfield  {title} {\bibinfo {title} {A simple formula for the average gate fidelity of a quantum dynamical operation},\ }\href {https://doi.org/10.1016/S0375-9601(02)01272-0} {\bibfield  {journal} {\bibinfo  {journal} {Physics Letters A}\ }\textbf {\bibinfo {volume} {303}},\ \bibinfo {pages} {249} (\bibinfo {year} {2002})}\BibitemShut {NoStop}%
\bibitem [{\citenamefont {Chalermpusitarak}\ \emph {et~al.}(2021)\citenamefont {Chalermpusitarak}, \citenamefont {Tonekaboni}, \citenamefont {Wang}, \citenamefont {Norris}, \citenamefont {Viola},\ and\ \citenamefont {{Paz-Silva}}}]{chalermpusitarakFrameBasedFilterFunctionFormalism2021}%
  \BibitemOpen
  \bibfield  {author} {\bibinfo {author} {\bibfnamefont {T.}~\bibnamefont {Chalermpusitarak}}, \bibinfo {author} {\bibfnamefont {B.}~\bibnamefont {Tonekaboni}}, \bibinfo {author} {\bibfnamefont {Y.}~\bibnamefont {Wang}}, \bibinfo {author} {\bibfnamefont {L.~M.}\ \bibnamefont {Norris}}, \bibinfo {author} {\bibfnamefont {L.}~\bibnamefont {Viola}},\ and\ \bibinfo {author} {\bibfnamefont {G.~A.}\ \bibnamefont {{Paz-Silva}}},\ }\bibfield  {title} {\bibinfo {title} {Frame-{{Based Filter-Function Formalism}} for {{Quantum Characterization}} and {{Control}}},\ }\href {https://doi.org/10.1103/PRXQuantum.2.030315} {\bibfield  {journal} {\bibinfo  {journal} {PRX Quantum}\ }\textbf {\bibinfo {volume} {2}},\ \bibinfo {pages} {030315} (\bibinfo {year} {2021})}\BibitemShut {NoStop}%
\bibitem [{\citenamefont {{DeepMind}}\ \emph {et~al.}(2020)\citenamefont {{DeepMind}}, \citenamefont {Babuschkin}, \citenamefont {Baumli}, \citenamefont {Bell}, \citenamefont {Bhupatiraju}, \citenamefont {Bruce}, \citenamefont {Buchlovsky}, \citenamefont {Budden}, \citenamefont {Cai}, \citenamefont {Clark}, \citenamefont {Danihelka}, \citenamefont {Dedieu}, \citenamefont {Fantacci}, \citenamefont {Godwin}, \citenamefont {Jones}, \citenamefont {Hemsley}, \citenamefont {Hennigan}, \citenamefont {Hessel}, \citenamefont {Hou}, \citenamefont {Kapturowski}, \citenamefont {Keck}, \citenamefont {Kemaev}, \citenamefont {King}, \citenamefont {Kunesch}, \citenamefont {Martens}, \citenamefont {Merzic}, \citenamefont {Mikulik}, \citenamefont {Norman}, \citenamefont {Papamakarios}, \citenamefont {Quan}, \citenamefont {Ring}, \citenamefont {Ruiz}, \citenamefont {Sanchez}, \citenamefont {Sartran}, \citenamefont {Schneider}, \citenamefont {Sezener}, \citenamefont {Spencer}, \citenamefont {Srinivasan}, \citenamefont {Stanojevi{\'c}}, \citenamefont {Stokowiec}, \citenamefont {Wang}, \citenamefont {Zhou},\ and\ \citenamefont {Viola}}]{deepmind2020jax}%
  \BibitemOpen
  \bibfield  {author} {\bibinfo {author} {\bibnamefont {{DeepMind}}}, \bibinfo {author} {\bibfnamefont {I.}~\bibnamefont {Babuschkin}}, \bibinfo {author} {\bibfnamefont {K.}~\bibnamefont {Baumli}}, \bibinfo {author} {\bibfnamefont {A.}~\bibnamefont {Bell}}, \bibinfo {author} {\bibfnamefont {S.}~\bibnamefont {Bhupatiraju}}, \bibinfo {author} {\bibfnamefont {J.}~\bibnamefont {Bruce}}, \bibinfo {author} {\bibfnamefont {P.}~\bibnamefont {Buchlovsky}}, \bibinfo {author} {\bibfnamefont {D.}~\bibnamefont {Budden}}, \bibinfo {author} {\bibfnamefont {T.}~\bibnamefont {Cai}}, \bibinfo {author} {\bibfnamefont {A.}~\bibnamefont {Clark}}, \bibinfo {author} {\bibfnamefont {I.}~\bibnamefont {Danihelka}}, \bibinfo {author} {\bibfnamefont {A.}~\bibnamefont {Dedieu}}, \bibinfo {author} {\bibfnamefont {C.}~\bibnamefont {Fantacci}}, \bibinfo {author} {\bibfnamefont {J.}~\bibnamefont {Godwin}}, \bibinfo {author} {\bibfnamefont {C.}~\bibnamefont {Jones}}, \bibinfo {author} {\bibfnamefont {R.}~\bibnamefont {Hemsley}}, \bibinfo {author} {\bibfnamefont {T.}~\bibnamefont {Hennigan}}, \bibinfo {author} {\bibfnamefont {M.}~\bibnamefont {Hessel}}, \bibinfo {author} {\bibfnamefont {S.}~\bibnamefont {Hou}}, \bibinfo {author} {\bibfnamefont {S.}~\bibnamefont {Kapturowski}}, \bibinfo {author} {\bibfnamefont {T.}~\bibnamefont {Keck}}, \bibinfo {author} {\bibfnamefont {I.}~\bibnamefont {Kemaev}}, \bibinfo {author} {\bibfnamefont {M.}~\bibnamefont {King}}, \bibinfo {author} {\bibfnamefont {M.}~\bibnamefont {Kunesch}}, \bibinfo {author} {\bibfnamefont {L.}~\bibnamefont {Martens}}, \bibinfo {author} {\bibfnamefont {H.}~\bibnamefont {Merzic}}, \bibinfo {author} {\bibfnamefont {V.}~\bibnamefont {Mikulik}}, \bibinfo {author} {\bibfnamefont {T.}~\bibnamefont {Norman}}, \bibinfo {author} {\bibfnamefont {G.}~\bibnamefont {Papamakarios}}, \bibinfo {author} {\bibfnamefont {J.}~\bibnamefont {Quan}}, \bibinfo {author} {\bibfnamefont {R.}~\bibnamefont {Ring}}, \bibinfo {author} {\bibfnamefont {F.}~\bibnamefont {Ruiz}}, \bibinfo {author} {\bibfnamefont {A.}~\bibnamefont {Sanchez}}, \bibinfo {author} {\bibfnamefont {L.}~\bibnamefont {Sartran}}, \bibinfo {author} {\bibfnamefont {R.}~\bibnamefont {Schneider}}, \bibinfo {author} {\bibfnamefont {E.}~\bibnamefont {Sezener}}, \bibinfo {author} {\bibfnamefont {S.}~\bibnamefont {Spencer}}, \bibinfo {author} {\bibfnamefont {S.}~\bibnamefont {Srinivasan}}, \bibinfo {author} {\bibfnamefont {M.}~\bibnamefont {Stanojevi{\'c}}}, \bibinfo {author} {\bibfnamefont {W.}~\bibnamefont {Stokowiec}}, \bibinfo {author} {\bibfnamefont {L.}~\bibnamefont {Wang}}, \bibinfo {author} {\bibfnamefont {G.}~\bibnamefont {Zhou}},\ and\ \bibinfo {author} {\bibfnamefont {F.}~\bibnamefont {Viola}},\ }\href@noop {} {\bibinfo {title} {The {{DeepMind JAX Ecosystem}}}} (\bibinfo {year} {2020})\BibitemShut {NoStop}%
\bibitem [{\citenamefont {Pathumsoot}(2025)}]{Pathumsoot_Inspeqtor_2025}%
  \BibitemOpen
  \bibfield  {author} {\bibinfo {author} {\bibfnamefont {P.}~\bibnamefont {Pathumsoot}},\ }\href {https://github.com/PorametPat/specq-stable} {\bibinfo {title} {{Inspeqtor}}} (\bibinfo {year} {2025})\BibitemShut {NoStop}%
\bibitem [{\citenamefont {AbuGhanem}(2025)}]{abughanemIBMQuantumComputers2025}%
  \BibitemOpen
  \bibfield  {author} {\bibinfo {author} {\bibfnamefont {M.}~\bibnamefont {AbuGhanem}},\ }\bibfield  {title} {\bibinfo {title} {{{IBM}} quantum computers: Evolution, performance, and future directions},\ }\href {https://doi.org/10.1007/s11227-025-07047-7} {\bibfield  {journal} {\bibinfo  {journal} {The Journal of Supercomputing}\ }\textbf {\bibinfo {volume} {81}},\ \bibinfo {pages} {687} (\bibinfo {year} {2025})}\BibitemShut {NoStop}%
\bibitem [{\citenamefont {Chow}\ \emph {et~al.}(2014)\citenamefont {Chow}, \citenamefont {Gambetta}, \citenamefont {Magesan}, \citenamefont {Abraham}, \citenamefont {Cross}, \citenamefont {Johnson}, \citenamefont {Masluk}, \citenamefont {Ryan}, \citenamefont {Smolin}, \citenamefont {Srinivasan},\ and\ \citenamefont {Steffen}}]{chowImplementingStrandScalable2014}%
  \BibitemOpen
  \bibfield  {author} {\bibinfo {author} {\bibfnamefont {J.~M.}\ \bibnamefont {Chow}}, \bibinfo {author} {\bibfnamefont {J.~M.}\ \bibnamefont {Gambetta}}, \bibinfo {author} {\bibfnamefont {E.}~\bibnamefont {Magesan}}, \bibinfo {author} {\bibfnamefont {D.~W.}\ \bibnamefont {Abraham}}, \bibinfo {author} {\bibfnamefont {A.~W.}\ \bibnamefont {Cross}}, \bibinfo {author} {\bibfnamefont {B.~R.}\ \bibnamefont {Johnson}}, \bibinfo {author} {\bibfnamefont {N.~A.}\ \bibnamefont {Masluk}}, \bibinfo {author} {\bibfnamefont {C.~A.}\ \bibnamefont {Ryan}}, \bibinfo {author} {\bibfnamefont {J.~A.}\ \bibnamefont {Smolin}}, \bibinfo {author} {\bibfnamefont {S.~J.}\ \bibnamefont {Srinivasan}},\ and\ \bibinfo {author} {\bibfnamefont {M.}~\bibnamefont {Steffen}},\ }\bibfield  {title} {\bibinfo {title} {Implementing a strand of a scalable fault-tolerant quantum computing fabric},\ }\href {https://doi.org/10.1038/ncomms5015} {\bibfield  {journal} {\bibinfo  {journal} {Nature Communications}\ }\textbf {\bibinfo {volume} {5}},\ \bibinfo {pages} {4015} (\bibinfo {year} {2014})}\BibitemShut {NoStop}%
\bibitem [{\citenamefont {{Javadi-Abhari}}\ \emph {et~al.}(2024)\citenamefont {{Javadi-Abhari}}, \citenamefont {Treinish}, \citenamefont {Krsulich}, \citenamefont {Wood}, \citenamefont {Lishman}, \citenamefont {Gacon}, \citenamefont {Martiel}, \citenamefont {Nation}, \citenamefont {Bishop}, \citenamefont {Cross}, \citenamefont {Johnson},\ and\ \citenamefont {Gambetta}}]{javadi-abhariQuantumComputingQiskit2024}%
  \BibitemOpen
  \bibfield  {author} {\bibinfo {author} {\bibfnamefont {A.}~\bibnamefont {{Javadi-Abhari}}}, \bibinfo {author} {\bibfnamefont {M.}~\bibnamefont {Treinish}}, \bibinfo {author} {\bibfnamefont {K.}~\bibnamefont {Krsulich}}, \bibinfo {author} {\bibfnamefont {C.~J.}\ \bibnamefont {Wood}}, \bibinfo {author} {\bibfnamefont {J.}~\bibnamefont {Lishman}}, \bibinfo {author} {\bibfnamefont {J.}~\bibnamefont {Gacon}}, \bibinfo {author} {\bibfnamefont {S.}~\bibnamefont {Martiel}}, \bibinfo {author} {\bibfnamefont {P.~D.}\ \bibnamefont {Nation}}, \bibinfo {author} {\bibfnamefont {L.~S.}\ \bibnamefont {Bishop}}, \bibinfo {author} {\bibfnamefont {A.~W.}\ \bibnamefont {Cross}}, \bibinfo {author} {\bibfnamefont {B.~R.}\ \bibnamefont {Johnson}},\ and\ \bibinfo {author} {\bibfnamefont {J.~M.}\ \bibnamefont {Gambetta}},\ }\href {https://doi.org/10.48550/ARXIV.2405.08810} {\bibinfo {title} {Quantum computing with {{Qiskit}}}} (\bibinfo {year} {2024})\BibitemShut {NoStop}%
\bibitem [{\citenamefont {McKay}\ \emph {et~al.}(2017)\citenamefont {McKay}, \citenamefont {Wood}, \citenamefont {Sheldon}, \citenamefont {Chow},\ and\ \citenamefont {Gambetta}}]{mckayEfficient$Z$Gates2017}%
  \BibitemOpen
  \bibfield  {author} {\bibinfo {author} {\bibfnamefont {D.~C.}\ \bibnamefont {McKay}}, \bibinfo {author} {\bibfnamefont {C.~J.}\ \bibnamefont {Wood}}, \bibinfo {author} {\bibfnamefont {S.}~\bibnamefont {Sheldon}}, \bibinfo {author} {\bibfnamefont {J.~M.}\ \bibnamefont {Chow}},\ and\ \bibinfo {author} {\bibfnamefont {J.~M.}\ \bibnamefont {Gambetta}},\ }\bibfield  {title} {\bibinfo {title} {Efficient \${{Z}}\$ gates for quantum computing},\ }\href {https://doi.org/10.1103/PhysRevA.96.022330} {\bibfield  {journal} {\bibinfo  {journal} {Physical Review A}\ }\textbf {\bibinfo {volume} {96}},\ \bibinfo {pages} {022330} (\bibinfo {year} {2017})}\BibitemShut {NoStop}%
\bibitem [{\citenamefont {Krantz}\ \emph {et~al.}(2019)\citenamefont {Krantz}, \citenamefont {Kjaergaard}, \citenamefont {Yan}, \citenamefont {Orlando}, \citenamefont {Gustavsson},\ and\ \citenamefont {Oliver}}]{krantzQuantumEngineersGuide2019}%
  \BibitemOpen
  \bibfield  {author} {\bibinfo {author} {\bibfnamefont {P.}~\bibnamefont {Krantz}}, \bibinfo {author} {\bibfnamefont {M.}~\bibnamefont {Kjaergaard}}, \bibinfo {author} {\bibfnamefont {F.}~\bibnamefont {Yan}}, \bibinfo {author} {\bibfnamefont {T.~P.}\ \bibnamefont {Orlando}}, \bibinfo {author} {\bibfnamefont {S.}~\bibnamefont {Gustavsson}},\ and\ \bibinfo {author} {\bibfnamefont {W.~D.}\ \bibnamefont {Oliver}},\ }\bibfield  {title} {\bibinfo {title} {A quantum engineer's guide to superconducting qubits},\ }\href {https://doi.org/10.1063/1.5089550} {\bibfield  {journal} {\bibinfo  {journal} {Applied Physics Reviews}\ }\textbf {\bibinfo {volume} {6}},\ \bibinfo {pages} {021318} (\bibinfo {year} {2019})}\BibitemShut {NoStop}%
\bibitem [{\citenamefont {Hyypp{\"a}}\ \emph {et~al.}(2024)\citenamefont {Hyypp{\"a}}, \citenamefont {Veps{\"a}l{\"a}inen}, \citenamefont {Papi{\v c}}, \citenamefont {Chan}, \citenamefont {Inel}, \citenamefont {Landra}, \citenamefont {Liu}, \citenamefont {Luus}, \citenamefont {Marxer}, \citenamefont {{Ockeloen-Korppi}}, \citenamefont {Orbell}, \citenamefont {Tarasinski},\ and\ \citenamefont {Heinsoo}}]{hyyppaReducingLeakageSingleQubit2024}%
  \BibitemOpen
  \bibfield  {author} {\bibinfo {author} {\bibfnamefont {E.}~\bibnamefont {Hyypp{\"a}}}, \bibinfo {author} {\bibfnamefont {A.}~\bibnamefont {Veps{\"a}l{\"a}inen}}, \bibinfo {author} {\bibfnamefont {M.}~\bibnamefont {Papi{\v c}}}, \bibinfo {author} {\bibfnamefont {C.~F.}\ \bibnamefont {Chan}}, \bibinfo {author} {\bibfnamefont {S.}~\bibnamefont {Inel}}, \bibinfo {author} {\bibfnamefont {A.}~\bibnamefont {Landra}}, \bibinfo {author} {\bibfnamefont {W.}~\bibnamefont {Liu}}, \bibinfo {author} {\bibfnamefont {J.}~\bibnamefont {Luus}}, \bibinfo {author} {\bibfnamefont {F.}~\bibnamefont {Marxer}}, \bibinfo {author} {\bibfnamefont {C.}~\bibnamefont {{Ockeloen-Korppi}}}, \bibinfo {author} {\bibfnamefont {S.}~\bibnamefont {Orbell}}, \bibinfo {author} {\bibfnamefont {B.}~\bibnamefont {Tarasinski}},\ and\ \bibinfo {author} {\bibfnamefont {J.}~\bibnamefont {Heinsoo}},\ }\bibfield  {title} {\bibinfo {title} {Reducing {{Leakage}} of {{Single-Qubit Gates}} for {{Superconducting Quantum Processors Using Analytical Control Pulse Envelopes}}},\ }\href {https://doi.org/10.1103/PRXQuantum.5.030353} {\bibfield  {journal} {\bibinfo  {journal} {PRX Quantum}\ }\textbf {\bibinfo {volume} {5}},\ \bibinfo {pages} {030353} (\bibinfo {year} {2024})}\BibitemShut {NoStop}%
\bibitem [{JAX(2018)}]{JAXComposableTransformations2018}%
  \BibitemOpen
  \href@noop {} {\bibinfo {title} {{{JAX}}: Composable transformations of {{Python}}+{{NumPy}} programs}} (\bibinfo {year} {2018})\BibitemShut {NoStop}%
\bibitem [{\citenamefont {{Jonathan Heek}}\ \emph {et~al.}(2024)\citenamefont {{Jonathan Heek}}, \citenamefont {{Anselm Levskaya}}, \citenamefont {{Avital Oliver}}, \citenamefont {{Marvin Ritter}}, \citenamefont {{Bertrand Rondepierre}}, \citenamefont {{Marvin Ritter}}, \citenamefont {{Andreas Steiner}},\ and\ \citenamefont {{Marc van Zee}}}]{jonathanheekFlaxNeuralNetwork2024}%
  \BibitemOpen
  \bibfield  {author} {\bibinfo {author} {\bibnamefont {{Jonathan Heek}}}, \bibinfo {author} {\bibnamefont {{Anselm Levskaya}}}, \bibinfo {author} {\bibnamefont {{Avital Oliver}}}, \bibinfo {author} {\bibnamefont {{Marvin Ritter}}}, \bibinfo {author} {\bibnamefont {{Bertrand Rondepierre}}}, \bibinfo {author} {\bibnamefont {{Marvin Ritter}}}, \bibinfo {author} {\bibnamefont {{Andreas Steiner}}},\ and\ \bibinfo {author} {\bibnamefont {{Marc van Zee}}},\ }\href@noop {} {\bibinfo {title} {Flax: {{A}} neural network library and ecosystem for {{JAX}}}} (\bibinfo {year} {2024})\BibitemShut {NoStop}%
\bibitem [{\citenamefont {Kidger}(2021)}]{diffrax}%
  \BibitemOpen
  \bibfield  {author} {\bibinfo {author} {\bibfnamefont {P.}~\bibnamefont {Kidger}},\ }\emph {\bibinfo {title} {On {{Neural Differential Equations}}}},\ \href@noop {} {Ph.D. thesis},\ \bibinfo  {school} {University of Oxford} (\bibinfo {year} {2021})\BibitemShut {NoStop}%
\bibitem [{\citenamefont {Bergholm}\ \emph {et~al.}(2022)\citenamefont {Bergholm}, \citenamefont {Izaac}, \citenamefont {Schuld}, \citenamefont {Gogolin}, \citenamefont {Ahmed}, \citenamefont {Ajith}, \citenamefont {Alam}, \citenamefont {{Alonso-Linaje}}, \citenamefont {AkashNarayanan}, \citenamefont {Asadi}, \citenamefont {Arrazola}, \citenamefont {Azad}, \citenamefont {Banning}, \citenamefont {Blank}, \citenamefont {Bromley}, \citenamefont {Cordier}, \citenamefont {Ceroni}, \citenamefont {Delgado}, \citenamefont {Matteo}, \citenamefont {Dusko}, \citenamefont {Garg}, \citenamefont {Guala}, \citenamefont {Hayes}, \citenamefont {Hill}, \citenamefont {Ijaz}, \citenamefont {Isacsson}, \citenamefont {Ittah}, \citenamefont {Jahangiri}, \citenamefont {Jain}, \citenamefont {Jiang}, \citenamefont {Khandelwal}, \citenamefont {Kottmann}, \citenamefont {Lang}, \citenamefont {Lee}, \citenamefont {Loke}, \citenamefont {Lowe}, \citenamefont {McKiernan}, \citenamefont {Meyer}, \citenamefont {{Monta{\~n}ez-Barrera}}, \citenamefont {Moyard}, \citenamefont {Niu}, \citenamefont {O'Riordan}, \citenamefont {Oud}, \citenamefont {Panigrahi}, \citenamefont {Park}, \citenamefont {Polatajko}, \citenamefont {Quesada}, \citenamefont {Roberts}, \citenamefont {S{\'a}}, \citenamefont {Schoch}, \citenamefont {Shi}, \citenamefont {Shu}, \citenamefont {Sim}, \citenamefont {Singh}, \citenamefont {Strandberg}, \citenamefont {Soni}, \citenamefont {Sz{\'a}va}, \citenamefont {Thabet}, \citenamefont {{Vargas-Hern{\'a}ndez}}, \citenamefont {Vincent}, \citenamefont {Vitucci}, \citenamefont {Weber}, \citenamefont {Wierichs}, \citenamefont {Wiersema}, \citenamefont {Willmann}, \citenamefont {Wong}, \citenamefont {Zhang},\ and\ \citenamefont {Killoran}}]{pennylane}%
  \BibitemOpen
  \bibfield  {author} {\bibinfo {author} {\bibfnamefont {V.}~\bibnamefont {Bergholm}}, \bibinfo {author} {\bibfnamefont {J.}~\bibnamefont {Izaac}}, \bibinfo {author} {\bibfnamefont {M.}~\bibnamefont {Schuld}}, \bibinfo {author} {\bibfnamefont {C.}~\bibnamefont {Gogolin}}, \bibinfo {author} {\bibfnamefont {S.}~\bibnamefont {Ahmed}}, \bibinfo {author} {\bibfnamefont {V.}~\bibnamefont {Ajith}}, \bibinfo {author} {\bibfnamefont {M.~S.}\ \bibnamefont {Alam}}, \bibinfo {author} {\bibfnamefont {G.}~\bibnamefont {{Alonso-Linaje}}}, \bibinfo {author} {\bibfnamefont {B.}~\bibnamefont {AkashNarayanan}}, \bibinfo {author} {\bibfnamefont {A.}~\bibnamefont {Asadi}}, \bibinfo {author} {\bibfnamefont {J.~M.}\ \bibnamefont {Arrazola}}, \bibinfo {author} {\bibfnamefont {U.}~\bibnamefont {Azad}}, \bibinfo {author} {\bibfnamefont {S.}~\bibnamefont {Banning}}, \bibinfo {author} {\bibfnamefont {C.}~\bibnamefont {Blank}}, \bibinfo {author} {\bibfnamefont {T.~R.}\ \bibnamefont {Bromley}}, \bibinfo {author} {\bibfnamefont {B.~A.}\ \bibnamefont {Cordier}}, \bibinfo {author} {\bibfnamefont {J.}~\bibnamefont {Ceroni}}, \bibinfo {author} {\bibfnamefont {A.}~\bibnamefont {Delgado}}, \bibinfo {author} {\bibfnamefont {O.~D.}\ \bibnamefont {Matteo}}, \bibinfo {author} {\bibfnamefont {A.}~\bibnamefont {Dusko}}, \bibinfo {author} {\bibfnamefont {T.}~\bibnamefont {Garg}}, \bibinfo {author} {\bibfnamefont {D.}~\bibnamefont {Guala}}, \bibinfo {author} {\bibfnamefont {A.}~\bibnamefont {Hayes}}, \bibinfo {author} {\bibfnamefont {R.}~\bibnamefont {Hill}}, \bibinfo {author} {\bibfnamefont {A.}~\bibnamefont {Ijaz}}, \bibinfo {author} {\bibfnamefont {T.}~\bibnamefont {Isacsson}}, \bibinfo {author} {\bibfnamefont {D.}~\bibnamefont {Ittah}}, \bibinfo {author} {\bibfnamefont {S.}~\bibnamefont {Jahangiri}}, \bibinfo {author} {\bibfnamefont {P.}~\bibnamefont {Jain}}, \bibinfo {author} {\bibfnamefont {E.}~\bibnamefont {Jiang}}, \bibinfo {author} {\bibfnamefont {A.}~\bibnamefont {Khandelwal}}, \bibinfo {author} {\bibfnamefont {K.}~\bibnamefont {Kottmann}}, \bibinfo {author} {\bibfnamefont {R.~A.}\ \bibnamefont {Lang}}, \bibinfo {author} {\bibfnamefont {C.}~\bibnamefont {Lee}}, \bibinfo {author} {\bibfnamefont {T.}~\bibnamefont {Loke}}, \bibinfo {author} {\bibfnamefont {A.}~\bibnamefont {Lowe}}, \bibinfo {author} {\bibfnamefont {K.}~\bibnamefont {McKiernan}}, \bibinfo {author} {\bibfnamefont {J.~J.}\ \bibnamefont {Meyer}}, \bibinfo {author} {\bibfnamefont {J.~A.}\ \bibnamefont {{Monta{\~n}ez-Barrera}}}, \bibinfo {author} {\bibfnamefont {R.}~\bibnamefont {Moyard}}, \bibinfo {author} {\bibfnamefont {Z.}~\bibnamefont {Niu}}, \bibinfo {author} {\bibfnamefont {L.~J.}\ \bibnamefont {O'Riordan}}, \bibinfo {author} {\bibfnamefont {S.}~\bibnamefont {Oud}}, \bibinfo {author} {\bibfnamefont {A.}~\bibnamefont {Panigrahi}}, \bibinfo {author} {\bibfnamefont {C.-Y.}\ \bibnamefont {Park}}, \bibinfo {author} {\bibfnamefont {D.}~\bibnamefont {Polatajko}}, \bibinfo {author} {\bibfnamefont {N.}~\bibnamefont {Quesada}}, \bibinfo {author} {\bibfnamefont {C.}~\bibnamefont {Roberts}}, \bibinfo {author} {\bibfnamefont {N.}~\bibnamefont {S{\'a}}}, \bibinfo {author} {\bibfnamefont {I.}~\bibnamefont {Schoch}}, \bibinfo {author} {\bibfnamefont {B.}~\bibnamefont {Shi}}, \bibinfo {author} {\bibfnamefont {S.}~\bibnamefont {Shu}}, \bibinfo {author} {\bibfnamefont {S.}~\bibnamefont {Sim}}, \bibinfo {author} {\bibfnamefont {A.}~\bibnamefont {Singh}}, \bibinfo {author} {\bibfnamefont {I.}~\bibnamefont {Strandberg}}, \bibinfo {author} {\bibfnamefont {J.}~\bibnamefont {Soni}}, \bibinfo {author} {\bibfnamefont {A.}~\bibnamefont {Sz{\'a}va}}, \bibinfo {author} {\bibfnamefont {S.}~\bibnamefont {Thabet}}, \bibinfo {author} {\bibfnamefont {R.~A.}\ \bibnamefont {{Vargas-Hern{\'a}ndez}}}, \bibinfo {author} {\bibfnamefont {T.}~\bibnamefont {Vincent}}, \bibinfo {author} {\bibfnamefont {N.}~\bibnamefont {Vitucci}}, \bibinfo {author} {\bibfnamefont {M.}~\bibnamefont {Weber}}, \bibinfo {author} {\bibfnamefont {D.}~\bibnamefont {Wierichs}}, \bibinfo {author} {\bibfnamefont {R.}~\bibnamefont {Wiersema}}, \bibinfo {author} {\bibfnamefont {M.}~\bibnamefont {Willmann}}, \bibinfo {author} {\bibfnamefont {V.}~\bibnamefont {Wong}}, \bibinfo {author} {\bibfnamefont {S.}~\bibnamefont {Zhang}},\ and\ \bibinfo {author} {\bibfnamefont {N.}~\bibnamefont {Killoran}},\ }\href@noop {} {\bibinfo {title} {{{PennyLane}}: {{Automatic}} differentiation of hybrid quantum-classical computations}} (\bibinfo {year} {2022}),\ \Eprint {https://arxiv.org/abs/1811.04968} {arXiv:1811.04968 [quant-ph]} \BibitemShut {NoStop}%
\bibitem [{\citenamefont {Liaw}\ \emph {et~al.}(2018)\citenamefont {Liaw}, \citenamefont {Liang}, \citenamefont {Nishihara}, \citenamefont {Moritz}, \citenamefont {Gonzalez},\ and\ \citenamefont {Stoica}}]{raytune}%
  \BibitemOpen
  \bibfield  {author} {\bibinfo {author} {\bibfnamefont {R.}~\bibnamefont {Liaw}}, \bibinfo {author} {\bibfnamefont {E.}~\bibnamefont {Liang}}, \bibinfo {author} {\bibfnamefont {R.}~\bibnamefont {Nishihara}}, \bibinfo {author} {\bibfnamefont {P.}~\bibnamefont {Moritz}}, \bibinfo {author} {\bibfnamefont {J.~E.}\ \bibnamefont {Gonzalez}},\ and\ \bibinfo {author} {\bibfnamefont {I.}~\bibnamefont {Stoica}},\ }\bibfield  {title} {\bibinfo {title} {Tune: A research platform for distributed model selection and training},\ }\href@noop {} {\bibfield  {journal} {\bibinfo  {journal} {arXiv preprint arXiv:1807.05118}\ } (\bibinfo {year} {2018})},\ \Eprint {https://arxiv.org/abs/1807.05118} {arXiv:1807.05118} \BibitemShut {NoStop}%
\end{thebibliography}%

\clearpage

\appendix

\section{Packages} \label{sec:packages}

We chose \texttt{jax} \cite{JAXComposableTransformations2018} as a Python library for manipulation of arrays. \texttt{jax} provides an automatic differentiation capability, allowing us to simplify implementation. Furthermore, the pseudo-random number generator in \texttt{jax} is highly reproducible. For the deep neural network model framework, we use \texttt{flax} \cite{jonathanheekFlaxNeuralNetwork2024} as it is a simple choice to implement the Blackbox model. On optimization, we use \texttt{optax} \cite{deepmind2020jax} for its flexibility. We implemented the Whitebox using \texttt{diffrax} \cite{diffrax}, testing it with \texttt{pennylane} \cite{pennylane}. For the hyperparameter tuning and distributing computation, we used \texttt{ray-tune} \cite{raytune}, while \texttt{ray-tune} uses \texttt{optuna} for hyperparameter tuning as we specified.

\section{Noise operator modelling} \label{sec:wo-modelling}
We model $W_O$ from \cref{eq:wo} predicted by the Blackbox as the following. Any Hermitian matrix, $A$, can be decomposed in terms of a unitary matrix, $Q$, and a diagonal matrix $D$ as $A = QDQ^\dagger$. The unitary matrix can be parametrized with three real parameters, $\alpha, \theta, \beta \in [0, 2 \pi ]$,
\begin{equation}
    Q (\alpha, \theta, \beta) = \begin{bmatrix}
        e^{i\alpha}\cos(\theta) & e^{i\beta}\sin(\theta)    \\
        e^{-i\beta}\sin(\theta) & -e^{-i\alpha}\cos(\theta)
    \end{bmatrix}.
\end{equation}
The diagonal matrix can be parametrized with two real parameters, $\lambda_1, \lambda_2 \in [-1, 1 ]$,
\begin{equation}
    D(\lambda_1, \lambda_2) = \begin{bmatrix}
        \lambda_1 & 0         \\
        0         & \lambda_2
    \end{bmatrix}.
\end{equation}
Thus, our deep neural network simply predicts 5 parameters for each Pauli observable, resulting in a prediction of 15 parameters given control variables.
\end{document}